\documentclass{aa}  

\usepackage{graphicx}
\usepackage{txfonts}
\usepackage{lipsum}
\usepackage{subcaption}        
\usepackage{lscape}               
\usepackage{placeins}             
\usepackage{rotating}
\usepackage{tabularx}
\usepackage[colorlinks,citecolor=blue]{hyperref}
\hypersetup{
    colorlinks = true,
    linkcolor = blue,
    anchorcolor = blue,
    citecolor = blue,
    filecolor = blue,
    urlcolor = blue
    }
\usepackage[normalem]{ulem}
\usepackage{siunitx}             
\usepackage{xcolor}              
\usepackage[utf8]{inputenc}      
\DeclareUnicodeCharacter{02BB}{\textokina}   
\newcommand{\textokina}{\raisebox{0.2ex}{\scalebox{0.8}{`}}}  
                                
\begin{document}

  \title{Cepheid metallicities from low-resolution near-infrared spectra: Validation against the optical reference scale}  
   \titlerunning{Cepheid metallicities from low-resolution near-infrared spectra}
   
   \author{M. A. Urbaneja\inst{1}\corrauth{miguel.urbaneja-perez@uibk.ac.at}        
        \and R. P. Kudritzki\inst{2,3}\email{kud@ifa.hawaii.edu}
        \and N. Przybilla\inst{1}\email{norbert.przybilla@uibk.ac.at}
}
        
   \institute{Universit\"at Innsbruck, Institut f\"ur Astro- und Teilchenphysik, Technikerstr. 25/8, 6020 Innsbruck, Austria
                 \and Institute for Astronomy, University of Hawai{\textokina}i~at Manoa, 2680 Woodlawn Drive, Honolulu, HI 96822, USA
                 \and LMU M\"unchen, Universit\"atssternwarte, Scheinerstr. 1, 81679 M\"unchen, Germany}

   \date{Received September 30, 20XX}

  \abstract
{Classical Cepheid metallicities trace recent chemical enrichment and help assess systematics in Cepheid-based distance measurements. They are derived mainly from high-resolution optical spectroscopy, which
defines the benchmark abundance scale but restricts measurements to nearby systems. Extending such studies to more distant Cepheids with extremely large telescopes requires complementary low-resolution
near-infrared approaches and careful treatment of model--data residuals.}
   {We tested whether low-resolution SpeX/IRTF $Y$+$J$-band spectra can deliver Cepheid metallicities on a scale compatible with homogenised high-resolution optical abundances, using Galactic calibrators as a first validation step towards more distant applications.} 
   {We analysed a homogeneous sample of 14 Galactic classical Cepheids observed with SpeX/IRTF, spanning periods from approximately 13 to 69 d, supplemented by two shorter-period Cepheids from the IRTF Spectral Library. The spectra were analysed with a Bayesian full-spectrum fitting method based on MARCS/TURBOSPECTRUM synthetic spectra and a regularised covariance likelihood, from which we inferred the global metallicity parameter $[{\rm M/H}]$. The resulting metallicities were compared with homogenised optical $[{\rm Fe/H}]$ measurements.}
   {For the primary 14-star homogeneous sample, the inferred metallicities reproduce the homogenised optical reference scale with a mean offset of $-0.05$ dex, a scatter of $0.09$ dex, and a robust scatter of $0.12$ dex. The median posterior uncertainty is $0.22$ dex. Including the two additional stars changes these values only slightly. Relative to simpler diagonal likelihoods, the use of a covariance-aware likelihood reduces the mean offset and improves the stability of the uncertainty calibration.}
   {Low-resolution $Y+J$ spectra contain usable metallicity information for classical Cepheids, provided that short-range correlated residuals and localised modelling imperfections are treated explicitly. The resulting SpeX-based near-infrared metallicity scale is tied to homogenised optical $[{\rm Fe/H}]$ measurements and provides a practical foundation for extending Cepheid metallicity work to larger and more distant samples.}

\keywords{stars: variables: Cepheids -- stars: abundances -- stars: atmospheres --
infrared: stars -- methods: data analysis -- methods: statistical}

\maketitle
\nolinenumbers 

\section{Introduction}
\label{sec:introduction}

Classical Cepheids are fundamental tracers of young stellar populations, and they play a central role in the extragalactic distance scale \citep[e.g.][]{Freedman2010,Riess2022}. Their metallicities provide probes of recent chemical enrichment, radial abundance gradients, and the composition of star-forming environments \citep[e.g.][]{Genovali2014,Luck2018,Trentin2024,Nunnari2026}. They are also important for assessing potential metallicity-related systematics in Cepheid-based distance measurements \citep[e.g.][]{Ripepi2020,Breuval2022}. Reliable Cepheid metallicities are therefore valuable both for studies of stellar populations and for distance-scale applications.

The most detailed abundance information for Galactic and nearby extragalactic Cepheids comes from high-resolution optical spectroscopy \citep[e.g.][]{Andrievsky2002,Kovtyukh2016,Lemasle2017,Romaniello2022}. Such analyses provide element-by-element abundances and define the reference metallicity scale against which complementary methods can be tested. However, optical spectroscopy becomes increasingly challenging for reddened Cepheids and for applications beyond the nearest systems, where extinction, crowding, and flux limitations restrict the accessible samples. Near-infrared spectroscopy offers a complementary approach because it is less affected by extinction and is well suited to cool, luminous stars \citep[e.g.][]{Ryde2009}. These advantages are particularly relevant for long-period Cepheids, which are among the brightest variables accessible in nearby galaxies and are therefore prime targets for future extragalactic metallicity studies, where Cepheid abundances can complement those derived from blue supergiants \citep[e.g.][]{Kudritzki2008, Kudritzki2024, Bresolin2025}.

Several studies have explored near-infrared abundance analyses of Cepheids and have shown that these spectra contain useful metallicity information \citep[e.g.][Nunnari et al. in prep]{Inno2019,Matsunaga2023,Catanzaro2026}. The specific question addressed here is whether lower-resolution, broad-wavelength SpeX/IRTF spectra can provide global metallicity estimates on a scale compatible with optical abundance measurements. At a resolving power of $R \approx 2000$, much of the metallicity information is distributed over blended atomic and molecular features rather than well-resolved, isolated  lines.

A Bayesian full-spectrum approach is well suited to extracting this information because it allows the stellar parameters, continuum terms, and instrumental broadening to be inferred simultaneously while propagating the corresponding uncertainties \citep[e.g.][]{Czekala2015,Casey2016,Ting2019}. At the same time, low-resolution near-infrared spectra exhibit correlated residuals arising from continuum placement, imperfect line lists, telluric correction, reduction effects, and limitations of the atmospheric models. In this regime, an explicit treatment of short-range correlated residuals provides a practical way to obtain stable parameter estimates while accounting for localised model-data mismatches.

In this paper, we validate this approach using low-resolution SpeX/IRTF $Y$- and $J$-band spectra of Galactic classical Cepheids. The inferred grid parameter is the global metallicity [M/H], defined in our models as a scaled-solar abundance pattern in which all metals are varied together relative to solar. We interpret this quantity as a global metallicity estimate and compare it with homogenised literature iron abundances [Fe/H]. This comparison ties the low-resolution near-infrared metallicity scale to the homogenised optical iron-abundance scale. 

The paper is organised as follows. In Sect.~\ref{sec:observations} we describe the SpeX observations, the extended comparison sample, and the literature reference metallicities. Section~\ref{sec:models_inference} presents the spectral models and Bayesian inference framework. In Sect.~\ref{sec:likelihood} we describe the adopted likelihood model. Section~\ref{sec:results} presents the metallicity validation. In Sect.~\ref{sec:likelihood_sensitivity} we discuss the sensitivity to the likelihood assumptions. In Sect.~\ref{sec:discussion} the results are placed in the broader context of Cepheid and young-star metallicity studies. Section~\ref{sec:conclusions} provides a summary of our conclusions.

\section{Observations and reference sample}
\label{sec:observations}

\subsection{Sample selection}
\label{subsec:sample_selection}
 
The sample analysed in this work consists of 16 Galactic classical Cepheids observed in the near-infrared with SpeX/IRTF. Fourteen of the stars were observed during a dedicated observing programme on 2 October 2014 and were selected to provide a homogeneous sample of bright classical Cepheids spanning periods of approximately 13--69 d. This range samples the luminous Cepheid regime most relevant for future applications to more distant systems while retaining spectra of sufficient quality for full-spectrum modelling at moderate resolving power.

We complemented the sample by two shorter-period classical Cepheids from the IRTF Spectral Library \citep{Rayner2009}, with periods of 5.4 and 2.5 d. Although these spectra were obtained independently of the main observing programme, they extend the period range explored in this study and were analysed using the same methodology.

A single near-infrared spectrum is analysed for each Cepheid. We did not attempt to model phase-dependent changes in the atmospheric structure. The present study therefore assesses the metallicity information recoverable from single-epoch spectra rather than performing a phase-resolved abundance analysis.

Unless explicitly stated otherwise, the primary validation presented throughout this paper is based on the homogeneous 14-star SpeX sample. The two additional Cepheids from the IRTF Spectral Library were used as a supplementary consistency check to assess the robustness of the inferred metallicity scale and to extend the explored period baseline.

\subsection{Observations}
\label{subsec:observations}

The observations were obtained with the SpeX spectrograph in short cross-dispersed (SXD) mode, which provides simultaneous wavelength coverage from approximately 0.8 to 2.5~$\mu{\rm m}$. In the present work, however, we restricted the analysis to the $Y$- and $J$-band regions between approximately 1.0 and 1.35~$\mu{\rm m}$. These wavelength intervals contain numerous atomic and molecular features that are sensitive to the atmospheric parameters and global metallicity.

A slit width of $0\farcs3$ was used, corresponding to a nominal resolving power of $R\sim2000$. The spectral resolution element was sampled by approximately two detector pixels. The observations followed standard near-infrared procedures, including nodding along the slit for sky subtraction. Telluric standard stars of spectral type A0V were observed close in time and airmass to the science targets to correct for atmospheric absorption.

The basic properties of the sample and the reference metallicities adopted for the external comparison are summarised in Table~\ref{tab:sample_observations}. The spectra have high signal-to-noise ratios, with a typical median $S/N \gtrsim 200$ in the fitted $Y$+$J$ wavelength region.

\subsection{Data reduction}
\label{subsec:data_reduction}

The data were reduced with the Spextool pipeline \citep{Cushing2004}, following the standard procedures for SpeX short cross-dispersed observations. The reduction includes flat-field correction, wavelength calibration, extraction of one-dimensional spectra, and the combination of the nodded exposures.

Telluric correction was performed using the observed A0V standard stars and the methodology described by \citet{Vacca2003}. This procedure removes most atmospheric absorption features, although residuals can remain in regions of strong telluric contamination and near hydrogen lines intrinsic to the telluric standards. Such regions were either excluded from the fit or assigned additional tolerance in the likelihood analysis described in Sect.~\ref{sec:likelihood} and Appendix~\ref{app:likelihood_tests}.

The final reduced spectra are provided with associated uncertainty
estimates from the reduction process, which are used directly in the statistical analysis. No additional smoothing was applied before the modelling. Any resampling required to match the model and observed wavelength grids was performed within the spectral-analysis workflow described in Sect.~\ref{subsec:forward_model}. Representative examples of the observed spectra are presented in Appendix~D.

\subsection{Optical reference metallicities}
\label{subsec:reference_metallicities}

The external comparison is based on optical literature iron abundances for the same Cepheids. The quantity inferred from the near-infrared spectra is the global metallicity parameter $[\mathrm{M/H}]$ of the model grid, which we compare with homogenised literature $[\mathrm{Fe/H}]$ values. The published abundances were placed on a common reference scale before the comparison. The adopted reference metallicity, its uncertainty, and the corresponding literature sources are listed in Table~\ref{tab:sample_observations}. The homogenisation procedure is described in Appendix~\ref{app:literature_homogenisation}.

\begin{table*}
\caption{Basic properties of the Cepheid sample.}
\label{tab:sample_observations}
\centering
\begin{tabular}{lrrrrc}
\hline
\multicolumn{1}{c}{Star} &
\multicolumn{1}{c}{Gaia DR3 source ID} &
\multicolumn{1}{c}{Pulsation period} &
\multicolumn{1}{c}{SpeX MJD} &
\multicolumn{1}{c}{$[\mathrm{Fe/H}]_{\rm hom}$} &
\multicolumn{1}{c}{Reference} \\
&
&
\multicolumn{1}{c}{(d)} &
\multicolumn{1}{c}{(d)} &
\multicolumn{1}{c}{(dex)} &
\\
\hline
S Vul        & 2027971514401523456 & 69.467417 & 56932.263 &    0.06$\pm$0.10 & [1] \\
GY Sge       & 1825211056159949440 & 51.566225 & 56932.243 &    0.22$\pm$0.06 & [2] \\
SV Vul       & 2027951173435143680 & 44.913759 & 56932.281 &    0.08$\pm$0.14 & [1] \\
V609 Cyg     & 2176050304426499840 & 31.072488 & 56932.305 &    0.06$\pm$0.17 & [2] \\
VX Cyg       & 1873250780732545920 & 20.133705 & 56932.460 &    0.08$\pm$0.10 & [3] \\
KX Cyg       & 2067549118302850304 & 20.053370 & 56932.347 &    0.05$\pm$0.12 & [2] \\
DR Cep       & 2193449320015954688 & 19.027248 & 56932.409 & $-$0.23$\pm$0.15 & [2] \\
CD Cyg       & 2058374144759464064 & 17.068686 & 56932.323 &    0.13$\pm$0.16 & [1] \\
RW Cam       &  473043922712140928 & 16.417580 & 56932.603 &    0.06$\pm$0.18 & [1] \\
SZ Cyg       & 2071433765909167232 & 15.114398 & 56932.375 & $-$0.03$\pm$0.10 & [4] \\
CH Cas       & 2016581673416790656 & 15.087595 & 56932.500 & $-$0.10$\pm$0.10 & [3] \\
RW Cas       &  508915489570374272 & 14.788901 & 56932.560 &    0.09$\pm$0.10 & [4] \\
TX Cyg       & 2161786374436607616 & 14.712201 & 56932.487 &    0.06$\pm$0.10 & [3] \\
GX Sge       & 1825437177598003456 & 12.895141 & 56932.203 &    0.20$\pm$0.14 & [2] \\
$\delta$ Cep & 2200153454733285248 &  5.366154 &        ... &   0.05$\pm$0.05 & [1] \\
DT Cyg       & 1853025642297186688 &  2.4992\tablefootmark{a} &        ... & $-$0.01$\pm$0.09 & [1] \\
\hline
\end{tabular}
\tablebib{
(1)~\citet[][]{Luck2018};
(2)~\citet[][]{Luck2011};
(3)~\citet[][]{Andrievsky2013};
(4)~\citet[][]{Bhardwaj2023}
}
\tablefoot{
\tablefoottext{a}{Value taken from the literature because no pulsation period is available in Gaia DR3.}
}
\end{table*}

\section{Spectral models and Bayesian inference}
\label{sec:models_inference}

The observed spectra were analysed with a Bayesian full-spectrum fitting approach based on the MAUI framework \citep{Urbaneja2026}. A grid of synthetic near-infrared spectra was used to train a statistical emulator that predicts the model spectrum as a function of the atmospheric parameters. The aim of the inference is to determine the stellar parameters that best reproduce the SpeX $Y$+$J$ spectra within a forward-modelling framework.

In this section, we describe the spectral grid, the emulator-based
forward model, the fitted parameters, the adopted priors, and the
posterior inference procedure. The adopted spectral grid and inference
setup, including the priors, are summarised in
Table~\ref{tab:model_inference_setup}. The likelihood model used to compare
the predicted spectra with the data is described separately in
Sect.~\ref{sec:likelihood}, while additional details of the synthetic
spectral calculations and model assumptions are provided in
Appendix~\ref{app:stellar_models}.

\subsection{Synthetic spectral grid}
\label{subsec:synthetic_grid}

The synthetic spectra were computed from one-dimensional hydrostatic MARCS model atmospheres \citep{Gustafsson2008} using the TURBOSPECTRUM radiative-transfer code \citep{Alvarez1998,Plez2012}. The grid was constructed to cover the range of atmospheric parameters expected for Galactic classical Cepheids. The four atmospheric parameters are the effective temperature, $T_{\rm eff}$; the surface gravity, $\log g$ (in cgs units); the global metallicity, $[\mathrm{M/H}]$; and the microturbulent velocity, $\xi$. These parameters define the atmospheric parameter vector,
\[
\boldsymbol{\theta}_{\rm atm}
=
\left(
T_{\rm eff},
\log g,
[\mathrm{M/H}],
\xi
\right).
\]
The adopted parameter ranges are
\[
\begin{aligned}
4500 &\leq T_{\rm eff} \leq 6750~{\rm K}, \\
0.0 &\leq \log g \leq 3.0~{\rm dex}, \\
-0.75 &\leq [{\rm M/H}] \leq +0.75~{\rm dex}, \\
1 &\leq \xi \leq 7~{\rm km\,s^{-1}}.
\end{aligned}
\]
The synthetic spectra were computed on a regular MARCS grid spanning these four parameter intervals. Further details on the model calculations and grid construction are provided in Appendix~\ref{app:stellar_models}. The metallicity parameter $[{\rm M/H}]$ represents a global scaling of the heavy-element content of the model atmosphere 
and should therefore be interpreted as a global metallicity parameter rather than as a line-by-line iron abundance. In the validation analysis presented below, the inferred $[{\rm M/H}]$ values are compared with homogenised optical literature $[{\rm Fe/H}]$ measurements to assess the consistency of the near-infrared metallicity scale with the optical reference scale.

The synthetic MARCS/TURBOSPECTRUM spectra were used to train a statistical emulator following the MAUI framework \citep{Urbaneja2026}. The spectra are first represented in a reduced basis using principal component analysis, after which the dependence of the retained coefficients on $T_{\rm eff}$, $\log g$, $[{\rm M/H}]$, and $\xi$ is learned from the model grid. During the inference, trial spectra are predicted by the trained emulator within the domain covered by this grid. Extrapolation beyond this domain is explicitly forbidden.

The analysis is restricted to selected wavelength intervals within the $Y$- and $J$-bands, chosen to include diagnostic regions while avoiding the most strongly contaminated telluric regions. The fitted windows and masks were defined before the final inference and kept unchanged throughout the production analysis, as described in Sect.~\ref{subsec:flagged_pixels}.

\subsection{Forward model}
\label{subsec:forward_model}

At each trial point, the emulator predicts the normalised synthetic spectrum corresponding to the proposed atmospheric parameters,
\[
S_\lambda(\boldsymbol{\theta}_{\rm atm})
=
S_\lambda
\left(
T_{\rm eff},
\log g,
[{\rm M/H}],
\xi
\right).
\]
This spectrum is then convolved with an instrumental kernel, resampled onto the observed wavelength grid, and adjusted by a local continuum correction before evaluation of the likelihood. For a wavelength window $j$, the model flux at observed pixel $i$ can be written schematically as
\begin{equation}
m_{ij}(\boldsymbol{\Theta})
=
C_j(\lambda_{ij};\boldsymbol{c}_j)
\left[
{\cal I}_j
\left\{
{\cal K}(R_{\rm eff})
\otimes
S(\boldsymbol{\theta}_{\rm atm})
\right\}
\right]_i ,
\end{equation}
where ${\cal K}(R_{\rm eff})$ is the instrumental line-spread function, $\otimes$ denotes convolution, ${\cal I}_j$ represents the adopted Doppler shift to the observed-frame wavelength scale and the subsequent resampling onto the observed pixels in window $j$, and $C_j(\lambda;\boldsymbol{c}_j)$ is a low-order multiplicative continuum correction described by Legendre polynomials defined over that window.

The implementation allows for an additional multiplicative telluric-transmission component, but this option was not used in the present analysis. Residual telluric imperfections are instead handled through masking, local flagged-pixel tolerance, and the covariance-aware likelihood described in Sect.~\ref{sec:likelihood}.

The nominal resolving power of the SpeX observations is approximately $R \simeq 2000$, but small differences may arise from observing conditions, extraction, wavelength calibration, and slit illumination. We therefore included a multiplicative resolving-power scale factor, $s_R$, such that the effective resolving power entering the spectral convolution is
\[
R_{\rm eff}
=
s_R\,R_{\rm nom}.
\]
This parameter accounts for modest deviations from the nominal resolution without requiring a separate empirical resolution estimate for each spectrum. At the spectral resolution of the present data, we did not include separate free parameters for rotational or macroturbulent broadening. Classical Cepheids are generally slow rotators, with projected rotational velocities of only a few kilometres per second  
\citep[e.g.][]{Bersier1996,Borra2017}. Cepheids also exhibit phase-dependent line broadening associated with pulsation and atmospheric velocity gradients. However, these velocity scales are small compared with the instrumental width at $R \simeq 2000$. Any residual star-to-star differences of this kind are therefore expected to be subdominant relative to the instrumental line-spread function and the residual treatment adopted in the likelihood.

Continuum placement was treated within the fitting procedure rather than by fixing a single externally normalised spectrum. In practice, each fitted wavelength window was assigned an independent low-order multiplicative continuum correction,
\[
C_j(\lambda;\boldsymbol{c}_j)
=
\sum_{k=0}^{n_c}
c_{jk}\,
P_k\!\left[x_j(\lambda)\right],
\]
where $P_k$ is the Legendre polynomial of order $k$ and $x_j(\lambda)$ is a normalised wavelength coordinate within the window. The continuum coefficients were not sampled as global Markov Chain Monte Carlo (MCMC) parameters. Instead, for each trial atmospheric model and wavelength window, they were re-estimated as local linear nuisance parameters before evaluating the likelihood. This approach reduces sensitivity to residual continuum and flux-calibration mismatches while preserving the local spectral information that constrains the atmospheric parameters.

The sampled global parameter vector can therefore be written schematically as
\[
\boldsymbol{\Theta}
=
\left(
T_{\rm eff},
\log g,
[{\rm M/H}],
\xi,
s_R
\right).
\]
The likelihood also includes the additional residual-error treatment described in Sect.~\ref{sec:likelihood}; those terms are not expanded here in order to keep the present section focused on the deterministic spectral model.

\subsection{Priors}
\label{subsec:priors}

Uniform priors were adopted for $T_{\rm eff}$, $[{\rm M/H}]$, and microturbulent velocity within the domain covered by the emulator training grid. Surface gravity was assigned an informative Gaussian prior based on the Cepheid period, reflecting the expected relation between period and flux-weighted gravity. This prior is useful because the $Y$+$J$ spectra at $R \simeq 2000$ provide only limited independent constraints on $\log g$.

For the resolving-power scale factor, we adopted a Gaussian prior centred on the nominal value,
\[
s_R \sim \mathcal{N}(1,0.15^2),
\]
restricted to physically plausible values. This prior allows moderate deviations from the nominal resolving power while preventing the resolution parameter from compensating for broader modelling residuals.
\paragraph{Surface-gravity prior.}

The near-infrared spectra contain only limited independent information on the surface gravity of Cepheids at the resolution and wavelength range considered here. We therefore adopted an external, physically motivated prior on $\log g$ based on the relation between pulsation period and flux-weighted gravity. Following \citet{Kudritzki2003}, the flux-weighted gravity is defined as
\begin{equation}
\log g_{\rm F}
=
\log g
-
4\log\left(
\frac{T_{\rm eff}}{10^4~{\rm K}}
\right).
\label{eq:flux_weighted_gravity}
\end{equation}
A relation between pulsation period and $\log g_{\rm F}$ is expected because the pulsation period is governed by the stellar mean density and therefore by the underlying combination of stellar mass, radius, and luminosity. Such a relation is supported by both stellar evolutionary models and empirical samples of Galactic and Magellanic Cloud Cepheids \citep{Anderson2016,Groenewegen2020,Groenewegen2023}.

Specifically, we adopted the recent calibration of \citet{Groenewegen2023},
\begin{equation}
\log g_{\rm F}(P)
=
(-0.853 \pm 0.014)\log P
+
(3.442 \pm 0.017),
\label{eq:fg_period_relation}
\end{equation}
where $P$ is the fundamental-mode pulsation period in days. This
relation has an RMS scatter of 0.059 dex. For comparison, \citet{Groenewegen2020} derived for Galactic Cepheids
\[
\log g_{\rm F}(P)
=
(-0.80 \pm 0.03)\log P
+
(3.43 \pm 0.03),
\]
with an RMS scatter of 0.16 dex. The broad agreement between the Galactic and Magellanic Cloud calibrations supports the use of the period--$\log g_{\rm F}$ relation as an external constraint on the spectroscopic fit.

At each step of the inference, Eq.~(\ref{eq:fg_period_relation}) was evaluated at the observed pulsation period and converted into a prior expectation for $\log g$ using the current sampled value of $T_{\rm eff}$. The surface-gravity prior was applied as a Gaussian term centred
on this value, with a width of $\sigma_{\rm prior}=0.20$ dex, which is substantially broader than the intrinsic scatter of the
empirical period--$\log g_{\rm F}$ relation. This deliberately conservative choice accounts for phase-dependent atmospheric
variations in single-epoch spectra, possible differences between the Galactic and Magellanic Cloud samples, and systematic
differences among spectroscopic gravity scales. The prior therefore does not fix the surface gravity, but regularises it towards the physically plausible 
range expected for a classical Cepheid of the observed period and prevents the fit from drifting towards gravity values that are not constrained by 
the SpeX $Y$+$J$ spectrum itself.

\begin{table*}
\caption{Spectral grid and inference setup.}
\label{tab:model_inference_setup}
\centering
\begin{tabular}{ll}
\hline\hline
\multicolumn{1}{c}{Quantity} &
\multicolumn{1}{c}{Adopted setup} \\
\hline
Atmosphere models & MARCS \\
Spectral synthesis & TURBOSPECTRUM \\
Emulator framework & MAUI, PCA-based reduced spectral representation \\
Wavelength range & Selected $Y$+$J$ windows, $\simeq 1.0$--$1.33~\mu{\rm m}$ \\
\hline
$T_{\rm eff}$ & $4500$--$6750~{\rm K}$ \\
$\log g$ & $0.0$--$3.0$ dex \\
$[{\rm M/H}]$ & $-0.75$--$+0.75$ dex \\
$\xi$ & $1$--$7~{\rm km\,s^{-1}}$ \\
\hline
Free stellar parameters & $T_{\rm eff}$, $\log g$, $[{\rm M/H}]$, $\xi$ \\
Instrumental parameter & Resolving-power scale factor $s_R$ \\
Continuum treatment & Local multiplicative low-order polynomials \\
Continuum parameters & Nuisance parameters fitted per wavelength window \\
\hline
Prior on $T_{\rm eff}$ & Uniform within grid boundaries \\
Prior on $[{\rm M/H}]$ & Uniform within grid boundaries \\
Prior on $\xi$ & Uniform within grid boundaries \\
Prior on $\log g$ & Gaussian period--$\log g_{\rm F}$ prior \\
Prior on $s_R$ & $\mathcal{N}(1,0.15^2)$, truncated to allowed range \\
\hline
Reported estimate & Posterior median \\
Reported uncertainty & Central 68\% credible interval \\
Validation quantity & SpeX-based $[{\rm M/H}]$ versus optical $[{\rm Fe/H}]$ \\
\hline
\end{tabular}
\tablefoot{The likelihood model, including the treatment of correlated residuals and locally problematic pixels, is described separately in Sect.~\ref{sec:likelihood}.}
\end{table*}

\subsection{Posterior inference}
\label{subsec:posterior_inference}

For each star, the posterior probability distribution was sampled independently. The posterior is proportional to the product of the likelihood and the priors,
\[
p(\boldsymbol{\Theta}\mid\mathbf{f}_{\rm obs})
\propto
\mathcal{L}
\left(
\mathbf{f}_{\rm obs}
\mid
\boldsymbol{\Theta}
\right)
\,p(\boldsymbol{\Theta}),
\]
where $\mathbf{f}_{\rm obs}$ denotes the observed spectrum. At each trial point, the deterministic model was evaluated and compared with the observed spectrum through the likelihood. The explicit form of $\mathcal{L}$, including the treatment of correlated residuals and locally problematic pixels, is given in Sect.~\ref{sec:likelihood}.

The reported stellar parameters were obtained from the marginal posterior distributions. Unless stated otherwise, we quote the posterior median as the point estimate and the central 68\% credible interval as the uncertainty.

\section{Likelihood model}
\label{sec:likelihood}

\subsection{Motivation}
\label{subsec:likelihood_motivation}

The likelihood defines how the observed SpeX spectra are compared with the forward-model spectra described in Sect.~\ref{sec:models_inference}. For a given set of atmospheric and nuisance parameters, we define the residual vector in fitted window $\jmath$ as
\begin{equation}
\mathbf{r}_j(\boldsymbol{\Theta})
=
\mathbf{f}_{{\rm obs},j}
-
\mathbf{m}_j(\boldsymbol{\Theta}),
\label{eq:residual_vector}
\end{equation}
where $\mathbf{f}_{{\rm obs},j}$ is the observed flux vector and $\mathbf{m}_j(\boldsymbol{\Theta})$ is the corresponding forward-model prediction evaluated on the observed wavelength grid. The latter includes the continuum treatment, spectral broadening, and interpolation described in Sect.~\ref{sec:models_inference}.

A purely diagonal likelihood provides a useful reference case, but it is not fully adequate for the final analysis. The residuals show correlated structure over a few pixels, as expected for low-resolution spectra affected by instrumental sampling, continuum adjustment, reduction residuals, imperfect line data, telluric correction, and limitations of the synthetic spectra; similar issues have been noted in empirical analyses of SpeX spectra \citep[e.g.][]{Villaume2017}. Residual autocorrelation diagnostics of our Cepheid fits indicate typical correlation lengths of approximately 3--5 pixels. At the present resolving power, these short-range correlations are more plausibly associated with these effects than with resolved stellar line-profile broadening, as in high-resolution spectroscopy.

The production likelihood includes a short-range covariance component together with a local treatment of spectral regions that are particularly susceptible to modelling or reduction residuals. The likelihood is intentionally regularised to account for the dominant residual correlations without allowing the noise model to absorb metallicity-sensitive spectral information.

\subsection{Reference diagonal likelihood}
\label{subsec:diagonal_likelihood}

As a reference case, we considered a diagonal Gaussian likelihood in which the residuals are assumed to be independent between pixels. In this case, the covariance matrix is
\begin{equation}
\mathbf{C}_{\rm diag}
=
{\rm diag}(\sigma_i^2),
\end{equation}
where $\sigma_i$ is the adopted per-pixel uncertainty, including both the observational and emulator contributions. The corresponding log-likelihood is
\begin{equation}
\ln \mathcal{L}
=
-\frac{1}{2}
\left[
\mathbf{r}^{\rm T}
\mathbf{C}_{\rm diag}^{-1}
\mathbf{r}
+
\ln |\mathbf{C}_{\rm diag}|
+
N\ln(2\pi)
\right].
\end{equation}
This reference likelihood is useful for assessing the information content of the spectra and for comparison with more conservative treatments. However, because it treats neighbouring pixels as statistically independent, it does not account for the short-range correlated residuals visible in the fits. It is therefore not adopted for the final metallicity estimates.

\subsection{Regularised covariance likelihood}
\label{subsec:covariance_likelihood}

For the production analysis, we adopted a Gaussian likelihood with a covariance matrix of the form
\begin{equation}
\mathbf{C}
=
\mathbf{C}_{\rm diag}
+
\mathbf{C}_{\rm cov},
\end{equation}
where $\mathbf{C}_{\rm diag}$ contains the diagonal variance terms and $\mathbf{C}_{\rm cov}$ describes an additional short-range correlated residual component. For simplicity, the window index is omitted below. The log-likelihood is
\begin{equation}
\ln \mathcal{L}
=
-\frac{1}{2}
\left[
\mathbf{r}^{\rm T}
\mathbf{C}^{-1}
\mathbf{r}
+
\ln |\mathbf{C}|
+
N\ln(2\pi)
\right].
\end{equation}
The correlated component is described by a kernel applied to the per-pixel uncertainties. It is constructed on the observed wavelength grid, but the kernel distance is evaluated in velocity space rather than in pixel or wavelength units. For each pair of valid pixels $i$ and $j$, we defined the local velocity separation as

\begin{equation}
\Delta v_{ij}
=
c\,
\frac{\lambda_j-\lambda_i}
{(\lambda_i+\lambda_j)/2}.
\end{equation}
The off-diagonal covariance term is then written as
\begin{equation}
C_{{\rm cov},ij}
=
\sigma_i\sigma_j\,
s_{\rm cov}^2\,
k\!\left(
\frac{|\Delta v_{ij}|}
{\ell_{v,ij}}
\right),
\end{equation}
where $k$ is a dimensionless correlation kernel that specifies how the covariance between two pixels depends on their separation in velocity space, $s_{\rm cov}$ sets the amplitude of the correlated component, and $\ell_{v,ij}$ is the local velocity scale obtained by converting the sampled pixel-space correlation length, $\ell_{\rm cov}$, using the observed wavelength grid. This formulation makes the covariance approximately wavelength invariant, in the sense that the same nominal pixel scale corresponds to the appropriate local velocity scale across the fitted spectral window.

To avoid introducing spurious long-range correlations, the covariance matrix is additionally tapered beyond approximately one correlation length using a Hann window. This restricts the correlated component to local residual structure and prevents the covariance model from describing broad spectral variations. 

The covariance hyperparameters are regularised with informative truncated Gaussian priors rather than being allowed to vary freely over broad ranges. Specifically, we adopted
\begin{equation}
\ell_{\rm cov}
\sim
\mathcal{N}(4.0,1.0^2),
\qquad
2.0 < \ell_{\rm cov} < 7.0,
\end{equation}
and
\begin{equation}
\log s_{\rm cov}
\sim
\mathcal{N}(-1.0,0.3^2),
\qquad
-1.8 < \log s_{\rm cov} < -0.2.
\end{equation}
These priors are centred on the few-pixel residual autocorrelation scale and the empirically calibrated covariance amplitude. They allow for modest adjustment from star to star while preventing the covariance model from becoming sufficiently flexible to absorb metallicity-sensitive spectral structure.

\subsection{Treatment of locally problematic pixels}
\label{subsec:flagged_pixels}

Some wavelength regions show localised residuals that are not well described by the short-range covariance alone. These residuals can arise from imperfectly modelled spectral features, residual telluric or reduction artefacts, or other local discrepancies between the observed and synthetic spectra. We therefore use two fixed masks to distinguish between pixels that should be excluded entirely and pixels that should remain in the fit but receive additional local tolerance.

A small number of known non-stellar or otherwise unsuitable features are excluded entirely from the likelihood through a fixed exclusion mask. In particular, the diffuse interstellar band (DIB) near 1.318~$\mu{\rm m}$ \citep[e.g.][]{Geballe2011} -- which is by far the strongest DIB in the $Y+J$ bands \citep[e.g.][]{Ebenbichler22} -- is removed because it is of interstellar origin. These excluded pixels play no further role in the inference.

A separate flagged-pixel mask is applied to the remaining fitted pixels. Unlike the exclusion mask, these pixels remain part of the likelihood but are permitted to have an inflated local variance. The flagged-pixel mask was defined from the normalised residuals of an initial baseline analysis. For each wavelength pixel, we considered the distribution of the normalised residuals, $r_i/\sigma_i$, across the stellar sample and identified pixels that repeatedly showed deviations exceeding three times the formal uncertainty, that is,
\[
\left|\frac{r_i}{\sigma_i}\right| > 3, 
\]
in at least 30\% of the stars. These pixels were then assigned to a fixed mask, $m_{{\rm flag},i}$, before the final inference was performed. This procedure ensures that the additional tolerance is applied only to diagnostically motivated wavelength regions, rather than being adjusted freely during the fit.

The diagonal variance was modified as
\begin{equation}
\sigma_i^2
\rightarrow
\sigma_i^2
\left(
1
+
m_{{\rm flag},i}
a_{\rm flag}^2
\right),
\end{equation}
where $a_{\rm flag}$ is the relative extra-tolerance amplitude. Here $m_{{\rm flag},i}=0$ for normal pixels and $m_{{\rm flag},i}>0$ for flagged or downgradable pixels; values between zero and one allow fractional tolerance to be assigned. The case $a_{\rm flag}=0$ recovers the regularised global-covariance likelihood without additional flagged-pixel tolerance.

The extra-tolerance amplitude was included in the sampled parameter vector, together with the stellar parameters and the global-covariance hyperparameters defined above, and it was regularised through an informative truncated Gaussian prior,
\begin{equation}
\log a_{\rm flag}
\sim
\mathcal{N}(0.5,0.3^2),
\qquad
-2 < \log a_{\rm flag} < 1.
\end{equation}
The prior allows additional variance in the flagged regions while preventing the tolerance term from becoming a flexible residual model capable of absorbing broad 
astrophysical information.

This treatment prevents a small number of problematic pixels from dominating the posterior while retaining the metallicity information distributed across the fitted $Y$+$J$ spectral range. Unlike a highly flexible residual model, the additional tolerance is restricted to predefined, diagnostically motivated wavelength regions and is used together with the regularised global covariance component.

\subsection{Adopted production configuration}
\label{subsec:production_likelihood}

The final analysis adopts the regularised covariance likelihood with local flagged-pixel tolerance and a Mat\'ern-3/2 correlation kernel. The production configuration is summarised in Table~\ref{tab:likelihood_configuration}. It represents a conservative model that accounts for the dominant correlated and localised residual structure while limiting the noise model's ability to absorb metallicity-sensitive information. This likelihood is used for the final metallicities presented in Sect.~\ref{sec:results}.

The diagonal likelihood is retained as a reference case, while
alternative likelihood prescriptions are used only as sensitivity
tests. The principal formulations and their main characteristics are
summarised in Table~\ref{tab:likelihood_summary}, with further
implementation details provided in Appendix~\ref{app:likelihood_tests}.

\begin{table*}
\caption{Likelihood configuration adopted in the production analysis.}
\label{tab:likelihood_configuration}
\centering
\setlength{\tabcolsep}{1.5mm}
\begin{tabular}{lll}
\hline
\multicolumn{1}{c}{Component} &
\multicolumn{1}{c}{Adopted choice} &
\multicolumn{1}{c}{Purpose} \\
\hline
Reference likelihood & Diagonal Gaussian & Baseline comparison \\
Production kernel & Mat\'ern-3/2 & Short-range residual correlations \\
Correlation length &
$\ell_{\rm cov}$\,$\sim$\,$\mathcal{N}(4.0,1.0^2)$, truncated to 2\,$-$\,7 pixels &
Empirical autocorrelation scale \\
Covariance amplitude &
$\log s_{\rm cov}$\,$\sim$\,$\mathcal{N}(-1.0,0.3^2)$, truncated to $-$1.8\,$-$\,0.2 &
Scales per-pixel uncertainties \\
Taper &
Hann taper at $\sim1\,\ell_{\rm cov}$ &
Restricts covariance to nearby pixels \\
Problematic pixels & Flagged-pixel tolerance & Inflates local diagonal variance \\
Extra tolerance &
$\log a_{\rm flag}$\,$\sim$\,$\mathcal{N}(0.5,0.3^2)$, truncated to $-$2\,$-$\,1 &
Regularised local tolerance \\
Broad covariance hyperparameters & Not adopted & Avoids absorbing metallicity information \\
\hline
\end{tabular}
\end{table*}

\section{Results}
\label{sec:results}

We now present the results obtained with the production likelihood described in Sect.~\ref{sec:likelihood}. Unless
stated otherwise, the validation statistics reported in this section refer to the homogeneous 14-star SpeX sample, which
defines the primary validation set of this work. Results for the full 16-star comparison sample, including the two
additional IRTF Spectral Library Cepheids, are discussed where relevant as a robustness check.

\begin{figure}
\centering
\includegraphics[width=.96\columnwidth]{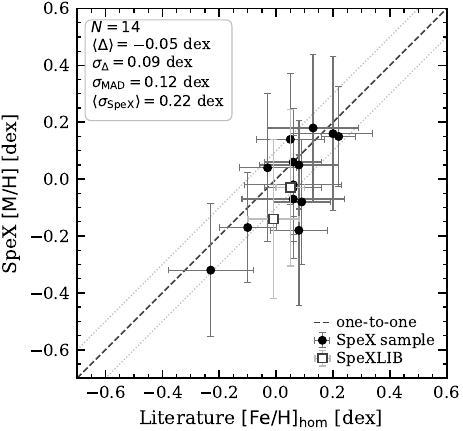}
\caption{SpeX-based global metallicities compared with homogenised
optical literature metallicities. Filled circles represent the 14
stars in the homogeneous SpeX sample, while open squares labelled
SpeXLIB represent the two comparison stars from the IRTF Spectral
Library. The dashed line indicates the one-to-one relation, and the
dotted lines indicate offsets of $\pm 0.1$ dex.}
\label{fig:mh_vs_feh}
\end{figure}

\subsection{Posterior constraints from the production analysis}
\label{subsec:posterior_constraints}

The production likelihood yields well-defined posterior constraints for the fitted parameters across the full sample. The covariance hyperparameters remain close to their regularising priors, with posterior medians of $\log s_{\rm cov}\simeq -1.0$, $\ell_{\rm cov}\simeq 4$ pixels, and $\log a_{\rm flag}\simeq 0.5$, as expected for the adopted residual model. The resolving-power scale factors likewise remain close to unity across the sample, indicating that no additional broadening component is required at a detectable level beyond modest deviations from the nominal instrumental resolution.

For the 14 homogeneous SpeX stars, the inferred effective temperatures span approximately 4600--6500 K, while the metallicities cover the range $[{\rm M/H}]\simeq -0.32$ to $+0.18$ dex. The two stars from the IRTF Spectral Library extend the period range to 5.37 and 2.50 d and yield metallicities of $[{\rm M/H}]=-0.03$ and $-0.14$ dex, respectively. The full posterior parameter table, including $T_{\rm eff}$, $\log g$, microturbulence, resolving-power scaling, and covariance hyperparameters, is provided in Appendix~\ref{app:production_parameters}.

\begin{table}
\caption{Comparison between the SpeX-based metallicities and literature values.}
\label{tab:metallicity_comparison}
\centering
\begin{tabular}{lrrr}
\hline
\multicolumn{1}{c}{Star} &
\multicolumn{1}{c}{$[{\rm M/H}]_{\rm SpeX}$} &
\multicolumn{1}{c}{$[{\rm Fe/H}]_{\rm hom}$} &
\multicolumn{1}{c}{$\Delta$} \\
&
\multicolumn{1}{c}{(dex)} &
\multicolumn{1}{c}{(dex)} &
\multicolumn{1}{c}{(dex)} \\
\hline
S  Vul   & $ 0.06^{+0.18}_{-0.20}$ & $ 0.06\pm0.10$ & $ 0.00$ \\
GY Sge   & $ 0.15^{+0.18}_{-0.17}$ & $ 0.22\pm0.06$ & $-0.07$ \\
SV Vul   & $ 0.05^{+0.15}_{-0.16}$ & $ 0.08\pm0.14$ & $-0.03$ \\
V609 Cyg & $-0.02^{+0.17}_{-0.18}$ & $ 0.06\pm0.17$ & $-0.08$ \\
VX Cyg   & $-0.18^{+0.25}_{-0.28}$ & $ 0.08\pm0.10$ & $-0.26$ \\
KX Cyg   & $ 0.14^{+0.24}_{-0.22}$ & $ 0.05\pm0.12$ & $ 0.09$ \\
DR Cep   & $-0.32^{+0.20}_{-0.27}$ & $-0.23\pm0.15$ & $-0.09$ \\
CD Cyg   & $ 0.18^{+0.30}_{-0.22}$ & $ 0.13\pm0.16$ & $ 0.05$ \\
RW Cam   & $-0.07^{+0.20}_{-0.22}$ & $ 0.06\pm0.18$ & $-0.13$ \\
SZ Cyg   & $ 0.04^{+0.24}_{-0.28}$ & $-0.03\pm0.10$ & $ 0.07$ \\
CH Cas   & $-0.17^{+0.19}_{-0.20}$ & $-0.10\pm0.10$ & $-0.07$ \\
RW Cas   & $-0.08^{+0.20}_{-0.24}$ & $ 0.09\pm0.10$ & $-0.17$ \\
TX Cyg   & $-0.03^{+0.18}_{-0.20}$ & $ 0.06\pm0.10$ & $-0.09$ \\
GX Sge   & $ 0.16^{+0.30}_{-0.24}$ & $ 0.20\pm0.14$ & $-0.04$ \\
$\delta$ Cep & $-0.03^{+0.30}_{-0.25}$ & $ 0.05\pm0.05$ & $-0.08$ \\
DT Cyg       & $-0.14^{+0.28}_{-0.28}$ & $-0.01\pm0.09$ & $-0.13$ \\
\hline
\end{tabular}
\tablefoot{The residuals are defined as
$\Delta = [{\rm M/H}]_{\rm SpeX} - [{\rm Fe/H}]_{\rm hom}$.}
\end{table}

The formal posterior uncertainties in $[{\rm M/H}]$ are typically of order 0.2--0.3 dex. These uncertainties include the effects of the covariance component and the local flagged-pixel tolerance and are therefore  
larger than those obtained from a purely diagonal likelihood.

\subsection{Comparison with homogenised optical metallicities}
\label{subsec:metallicity_validation}

Figure~\ref{fig:mh_vs_feh} compares the SpeX-based $[{\rm M/H}]$ estimates with the homogenised optical $[{\rm Fe/H}]$ values. The filled circles correspond to the 14 homogeneous SpeX targets that define the primary validation sample, while the two additional Cepheids adopted from the IRTF Spectral Library are shown separately to distinguish their provenance and to assess the robustness of the inferred metallicity scale. The corresponding star-by-star values are listed in Table~\ref{tab:metallicity_comparison}, together with the residuals used to compute the validation statistics.

\begin{figure*}
\centering
\includegraphics[width=.86\textwidth]{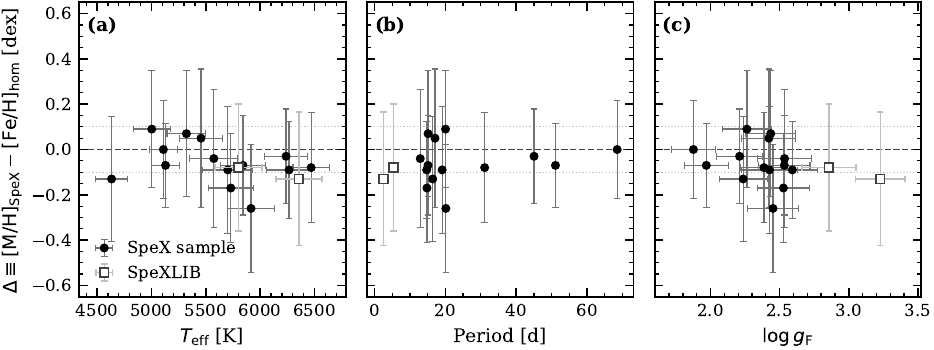}
\caption{Metallicity residuals, defined as
$\Delta=[{\rm M/H}]_{\rm SpeX}-[{\rm Fe/H}]_{\rm hom}$, as a function
of effective temperature, pulsation period, and flux-weighted gravity.
Filled circles represent the 14 stars in the homogeneous SpeX sample,
while open squares labelled SpeXLIB represent the two comparison stars
from the IRTF Spectral Library. The dashed horizontal line marks zero,
and the dotted lines indicate offsets of $\pm 0.1$ dex.}
\label{fig:residual_trends}
\end{figure*}

For the primary 14-star homogeneous SpeX sample, the mean offset relative to the homogenised optical scale is
\begin{equation}
\langle \Delta[{\rm M/H}] \rangle
=
-0.054~{\rm dex},
\end{equation}
where
\begin{equation}
\Delta[{\rm M/H}]
=
[{\rm M/H}]_{\rm SpeX}
-
[{\rm Fe/H}]_{\rm hom}.
\end{equation}
The standard deviation of the residuals is
\begin{equation}
\sigma_{\Delta}
=
0.085~{\rm dex},
\end{equation}
and the robust scatter $\sigma_{\rm MAD}$, defined as $1.4826$\,times the median absolute deviation (MAD), is
\begin{equation}
\sigma_{\rm MAD}
=
0.119~{\rm dex}.
\end{equation}

Including the two additional Cepheids from the IRTF Spectral Library changes the results only slightly, yielding a mean offset of $-0.061$ dex, a scatter of $0.082$ dex, and a robust scatter of $0.104$ dex. The primary 14-star validation sample therefore reproduces the homogenised optical reference scale at approximately the 0.1 dex level, and the full 16-star comparison sample supports the same conclusion.

The small negative mean offset is modest, given the differences in wavelength range, spectral resolution, and modelling assumptions. It might also reflect the fact that the near-infrared analysis constrains a global metallicity parameter, $[{\rm M/H}]$, whereas the optical reference values are iron abundances, $[{\rm Fe/H}]$, and the two quantities are not strictly identical.

\subsection{Residuals and uncertainty assessment}
\label{subsec:residuals_uncertainties}

We also assessed the residuals in units of the combined SpeX and literature uncertainties,
\begin{equation}
z
=
\frac{
[{\rm M/H}]_{\rm SpeX}
-
[{\rm Fe/H}]_{\rm hom}
}{
\sqrt{
\sigma_{\rm SpeX}^2
+
\sigma_{\rm hom}^2
}
}.
\end{equation}
For the primary 14-star homogeneous SpeX sample, the mean normalised residual is
\begin{equation}
\langle z \rangle
=
-0.22,
\end{equation}
and the root-mean-square normalised residual is
\begin{equation}
{\rm RMS}(z)
=
0.383.
\end{equation}

The median posterior uncertainty in $[{\rm M/H}]$ is 0.22 dex. The low RMS value of $z$ indicates that the combined uncertainties are conservative relative to the empirical star-to-star residuals. The posterior uncertainties should therefore be interpreted as conservative rather than as empirically calibrated one-sigma errors.

For the full 16-star comparison sample, the corresponding values are $\langle z\rangle=-0.238$ and ${\rm RMS}(z)=0.382$, again indicating that inclusion of the two additional stars does not materially change the uncertainty diagnostics.

\subsection{Trends with stellar parameters and period}
\label{subsec:residual_trends}

Figure~\ref{fig:residual_trends} shows the metallicity residuals as a function of effective temperature, pulsation period, and flux-weighted gravity. The effective temperatures and surface gravities used in this comparison are the posterior median values from the production analysis, while the periods are the adopted pulsation periods listed in Table~\ref{tab:sample_observations}. 

No obvious visual trend is present with any of these quantities. Within the present sample, the metallicity residuals do not show an evident dependence on effective temperature, period, or flux-weighted gravity. These diagnostics are necessarily qualitative given the modest sample size, but they suggest that the inferred metallicity scale behaves consistently across the parameter range explored here.

\subsection{Summary of production-likelihood performance}
\label{subsec:production_summary}

The principal validation metrics obtained with the adopted production likelihood are summarised in the preceding subsections. For the primary sample, the production likelihood gives a mean metallicity offset of $-0.054$ dex, a residual scatter of 0.085 dex, and a robust scatter of 0.119 dex relative to the homogenised optical scale. The median posterior uncertainty in $[{\rm M/H}]$ is 0.22 dex, and the RMS of the normalised residuals is 0.383. These statistics demonstrate consistency with the homogenised optical reference scale while maintaining conservative uncertainties.

Overall, the production likelihood reproduces the homogenised optical metallicity scale with a scatter of approximately 0.1 dex and no obvious residual trends with stellar parameters, supporting the use of low-resolution $Y+J$ spectroscopy as a practical tool for Cepheid metallicity studies.

\section{Sensitivity to likelihood assumptions}
\label{sec:likelihood_sensitivity}

\begin{table*}
\centering
\caption{Sensitivity of the metallicity validation to the adopted likelihood model.}
\label{tab:likelihood_sensitivity}
\begin{tabular}{lcccccc}
\hline
Likelihood variant & $\langle\Delta\rangle$ & $\sigma_\Delta$ & $\sigma_{\rm MAD}$ & median $\sigma_{[{\rm M/H}]}$ & $\langle z\rangle$ & RMS$(z)$ \\
& dex & dex & dex & dex & & \\
\hline
Diagonal Gaussian                         & $-0.102$ & 0.121 & 0.133 & 0.19 & $-0.492$ & 0.733 \\
Diagonal + robust flagged regions  & $-0.119$ & 0.071 & 0.089 & 0.24 & $-0.462$ & 0.539 \\
Wavelet/red-noise diagnostic          & $-0.097$ & 0.129 & 0.119 & 0.18 & $-0.496$ & 0.808 \\
Production likelihood                       & $-0.054$ & 0.085 & 0.119 & 0.22 & $-0.220$ & 0.383 \\
PCA low-rank diagnostic                 & $-0.066$ & 0.137 & 0.148 & 0.17 & $-0.367$ & 0.774 \\
Global covariance + PCA low-rank diagnostic & $-0.069$ & 0.129 & 0.133 & 0.17 & $-0.366$ & 0.708 \\
Global covariance + NMF diagnostic         & $-0.066$ & 0.111 & 0.089 & 0.21 & $-0.296$ & 0.532 \\
\hline
\end{tabular}
\end{table*}

The production results presented in Sect.~\ref{sec:results} were obtained with the regularised covariance likelihood described in Sect.~\ref{sec:likelihood}. Here we assess the sensitivity of the inferred metallicities to this choice. The goal is not to provide an exhaustive comparison of residual models but to verify that the adopted likelihood yields stable metallicities, conservative uncertainties, and a controlled description of the residual structure present in the low-resolution spectra.

Table~\ref{tab:likelihood_sensitivity} summarises the principal likelihood variants considered in this assessment. The statistics are intended as diagnostic quantities that illustrate the relative behaviour of different residual models. The primary validation remains based on the 14 homogeneous SpeX targets discussed in Sect.~\ref{sec:results}. The production-likelihood row adopts the final values from Sect.~\ref{sec:results}, while the remaining rows provide diagnostic comparisons with alternative residual treatments. Additional implementation details and exploratory likelihood variants are described in Appendix~\ref{app:likelihood_tests}.

\subsection{Reference likelihoods}

The diagonal Gaussian likelihood provides a useful reference case. It already recovers a metallicity scale broadly compatible with the optical reference values, demonstrating that the SpeX $Y$+$J$ spectra contain measurable metallicity information. However, it gives a larger mean offset and scatter than the production analysis, and its normalised residuals show a stronger systematic displacement. This behaviour is expected because the residuals exhibit short-range correlations that are not captured by a purely diagonal likelihood.

A more robust diagonal variant, in which diagnostically problematic regions are treated with a Student-$t$ likelihood, reduces the empirical scatter. This confirms that a limited number of wavelength regions can noticeably influence the fit. Nevertheless, this model remains diagonal outside those regions and therefore does not account for the short-range residual correlations seen across the fitted windows. We therefore regard the diagonal and robust-diagonal cases as useful baselines rather than final production choices.

The wavelet or red-noise diagnostic provides an alternative description of correlated residual structure. In the present application, however, it does not improve the validation statistics relative to the regularised covariance likelihood and is therefore retained only as a consistency check.

\subsection{Regularised covariance and local tolerance}

The covariance tests show that including a short-range correlated residual component improves the stability of the metallicity scale. With a Matérn-3/2 covariance term motivated by the residual autocorrelation scale, the mean offset is reduced relative to the diagonal reference case, and the residual scatter remains
close to 0.09 dex for the primary 14-star validation sample. This supports the interpretation that the dominant residual structure is not purely pixel-to-pixel noise but includes short-range correlations arising from the combined effects of spectral sampling, continuum adjustment, reduction residuals, and imperfections in the synthetic spectra.

The covariance term provides a more appropriate statistical description of the residuals at the resolution of the data. The regularising priors on $s_{\rm cov}$ and $\ell_{\rm cov}$ limit the flexibility of the covariance model and prevent it from absorbing metallicity-sensitive spectral information.

The tests also demonstrate the importance of treating localised problematic regions separately from the global covariance. A model with fixed covariance amplitude and length scale already improves the agreement with the optical metallicity scale relative to the diagonal reference case. Adding a local flagged-pixel tolerance further improves
the normalised-residual diagnostics while maintaining a small empirical scatter. This behaviour is consistent with the interpretation that most of the residual structure is short-range and broadly distributed, while a smaller subset of pixels requires additional local tolerance.

\subsection{Limits of more flexible residual models}

More flexible residual prescriptions were also explored as
diagnostics. When the covariance amplitude is allowed to vary over a
broad range, the posterior favours large covariance amplitudes and the
median metallicity uncertainty increases substantially. Although the RMS normalised residual then appears well behaved, this comes at the cost of broader posteriors and a larger metallicity offset and scatter. This suggests that an overly flexible covariance model can absorb spectral structure that would otherwise constrain the atmospheric parameters.  

This behaviour motivates the regularised production choice. At the spectral resolution considered here, the residual model must be flexible enough to account for correlated and localised imperfections but not so flexible that it removes the metallicity information distributed over blended $Y$- and $J$-band features. We therefore do not adopt broad free covariance amplitudes, nor more complex residual decompositions, for the production metallicities. Such models remain useful diagnostics, but their interpretation is limited by the modest sample size and by the fact that different sources of residual structure can be mixed together.

Overall, the sensitivity tests support the adopted production likelihood as a controlled compromise between an overly restrictive diagonal model and more flexible residual descriptions. The comparison with the optical reference scale remains stable when short-range correlated residuals and localised problematic pixels are treated in a regularised manner, while substantially more flexible residual models do not improve the external validation.

\section{Discussion}
\label{sec:discussion}

\subsection{Metallicity information in low-resolution near-infrared spectra}
\label{subsec:discussion_information}

The results presented in Sect.~\ref{sec:results} show that low-resolution SpeX spectra contain useful metallicity information for classical Cepheids, even when the analysis is restricted to the $Y$- and $J$-band regions and to a resolving power of approximately $R \simeq 2000$. This is not the regime of classical high-resolution abundance analysis, in which individual Fe lines can be measured and interpreted separately. Instead, the metallicity sensitivity is distributed across many blended atomic and molecular features, whose combined response is captured by the full-spectrum modelling approach.

The relevant information is likely carried by a combination of weak and moderately strong features across the fitted windows rather than by a small number of isolated diagnostics. This distributed information content is well suited to a Bayesian full-spectrum analysis, which can combine many low-contrast spectral signatures while marginalising over temperature, microturbulence, spectral resolution, and local continuum adjustments. At the same time, it makes the analysis sensitive to the statistical treatment of residuals, because imperfections in the models, continuum placement, telluric correction, and reduction affect groups of neighbouring pixels rather than isolated points.
The agreement with the homogenised optical scale indicates that these low-resolution near-infrared spectra provide a practical empirical metallicity estimator for Cepheids.

\subsection{Comparison with optical Cepheid metallicities}
\label{subsec:discussion_optical_comparison}

The comparison with optical literature metallicities provides the external validation of the near-infrared scale. The comparison presented in Sect.~~\ref{sec:results} shows that the SpeX-based metallicities are consistent with the homogenised optical reference scale at the level of approximately 0.1 dex. The
remaining residuals are modest and do not suggest any significant systematic discrepancy between the two abundance scales.

The comparison should nevertheless be interpreted in the context of the different quantities being inferred. The optical reference values are high-resolution iron abundances, whereas the present analysis constrains a global metallicity parameter from low-resolution near-infrared spectra. Therefore, the small negative offset found here does not indicate significant tension between the two scales but instead supports the interpretation of the SpeX metallicities as an externally anchored near-infrared metallicity scale.

The absence of an obvious residual trend with effective temperature, period, or flux-weighted gravity provides an additional consistency check. This is relevant because Cepheid spectra change substantially over the temperature range sampled here, and because the period range of the homogeneous sample spans the luminous Cepheid regime most relevant for future extragalactic applications. Within the present sample, the metallicity residuals do not indicate a clear bias associated with these quantities. This suggests that the inferred scale behaves consistently across the validation range. Nevertheless, the sample remains modest, and the trend tests should be viewed as qualitative diagnostics rather than a definitive exclusion of weak parameter-dependent systematics.

\subsection{Importance of residual modelling}
\label{subsec:discussion_residuals}

A central result of this work is that the quality of the metallicity validation depends not only on the spectral information present in the data but also on the statistical treatment of model-data residuals. The diagonal likelihood already recovers a broadly consistent metallicity scale, but the 
production analysis gives a more stable comparison with the optical reference scale and more conservative uncertainties. This behaviour is consistent with the residual structure seen in the spectra: at low resolution, neighbouring pixels are not fully independent, and localised wavelength regions can be affected by stronger residuals.

The adopted production likelihood therefore represents a balance between realism and regularisation. It accounts for correlated and localised imperfections while remaining sufficiently constrained for the metallicity information to remain identifiable.

This point is likely to be relevant beyond the present data set. Low-resolution near-infrared spectra of cool supergiants and Cepheids contain broad blends, molecular contributions, telluric residuals, and continuum-normalisation challenges. Reliable metallicity work in this regime therefore requires a residual model that is physically and empirically motivated but not so flexible that it removes the signal of interest. The present validation suggests that such a controlled treatment can place low-resolution near-infrared metallicities on a scale compatible with optical Cepheid abundances.

\subsection{Limitations and future applications}
\label{subsec:discussion_limitations_outlook}

Several limitations define the scope of the present validation. First, the primary validation sample contains only 14 homogeneous SpeX targets. This is sufficient for an initial external validation, but larger samples will be needed to test for weaker systematics with period, temperature, phase, metallicity, and data quality. Secondly, each star is represented by a single-epoch spectrum. Cepheids have phase-dependent atmospheric structures, and although metallicity should not vary with phase, the spectral diagnostics used to infer it can vary throughout the pulsation cycle. Testing phase dependence with multi-epoch observations will therefore be an important next step.

A further limitation is that the validation compares a near-infrared global metallicity parameter, $[{\rm M/H}]$, with optical iron abundances, $[{\rm Fe/H}]$. The good agreement found here supports the external anchoring of the near-infrared scale, but continued comparison with high-resolution optical studies remains important. The present analysis also depends on the adopted MARCS model atmospheres, TURBOSPECTRUM spectral synthesis, and the associated line lists, and improvements in near-infrared atomic and molecular data may therefore lead to refinements of the metallicity scale.

In its present form, the analysis is restricted to selected $Y$+$J$
windows. To assess how strongly the results depend on the available
wavelength coverage, we also performed a complementary $J$-band-only
analysis, presented in Appendix~\ref{app:jband}. This test recovers a broadly
consistent metallicity scale, but with a larger systematic offset and
slightly larger posterior uncertainties than the preferred combined
$Y$+$J$ analysis, reinforcing the advantage of the broader wavelength
coverage. Extending the approach further to the $H$- and $K$-bands may add useful information, but it will require renewed assessment of model-data residuals, telluric sensitivity, and possible wavelength-dependent systematics.

The broader motivation for this approach is the extension of direct 
Cepheid metallicity measurements to the Local Group and nearby galaxies. 
Near-infrared spectroscopy is naturally advantageous for reddened 
Galactic Cepheids, but its principal long-term value lies in reaching 
systems where high-resolution optical spectroscopy of large Cepheid 
samples remains impractical. A validated low-resolution near-infrared 
metallicity scale would provide a practical route to such measurements. 

High-multiplex instruments like the Multi-Object Optical and Near-infrared Spectrograph  \citep[MOONS,][]{Cirasuolo2020} on the Very Large Telescope (VLT) of the European Southern Observatory (ESO) will provide a natural way to test this approach on larger Cepheid samples in favourable nearby-galaxy fields in the near future. The Extremely Large Telescope (ELT) facilities are expected to extend the reach to fainter or more crowded systems. Diffraction-limited observations with the ELTs will only be feasible in the near-infrared because of limitations imposed by the adaptive optics systems. At the currently most advanced facility, both in scope and in advance of the construction progress -- the ESO ELT -- the multi-object spectrograph MOSAIC \citep{Hammer2021} will make Cepheid spectroscopy feasible beyond the Local Group in about a decade; at first light, only pilot studies of individual Cepheids will be feasible, such as with the spectroscopic mode of the Multi-AO Imaging Camera for Deep Observations \citep[][though at higher $R\simeq 20\,000$]{Sturm2024}.

This path is complementary to quantitative spectroscopy of blue 
supergiants, which has already been used extensively to determine 
metallicities and abundance gradients in Local Group and nearby galaxies 
\citep[e.g.][]{Urbaneja2008,U2009,Hosek2014,Berger2018}. Cepheids 
trace the same young stellar populations and serve as primary distance 
indicators, so a homogeneous near-infrared metallicity scale anchored 
to optical abundance measurements would provide an important bridge 
between the two approaches, enabling independent and complementary 
probes of the present-day chemical composition of nearby galaxies. The 
Large Magellanic Cloud offers a useful illustration: Cepheid and 
blue-supergiant analyses there yield consistent iron abundances 
\citep[e.g.][]{Lemasle2017,Urbaneja2017}.

The present work provides a first validation of a low-resolution near-infrared metallicity scale for Galactic classical Cepheids. Future applications should expand the calibration sample, test phase dependence, explore additional near-infrared wavelength ranges, and compare the resulting Cepheid metallicities with both high-resolution Cepheid abundances and blue-supergiant metallicity measurements in Local Group and nearby galaxies.

\section{Conclusions}
\label{sec:conclusions}

{We have tested whether low-resolution near-infrared spectra can provide reliable metallicity estimates for classical Cepheids. The analysis was based on $Y$- and $J$-band SpeX/IRTF spectra of 16 Galactic Cepheids. Fourteen stars form a homogeneous primary validation sample spanning periods from approximately 13 to 69 d, while two additional short-period Cepheids from the IRTF Spectral Library extend the explored period baseline.}

The spectra were analysed with a Bayesian full-spectrum method based on MARCS model atmospheres, TURBOSPECTRUM spectral synthesis, and the MAUI framework. The fitted metallicity parameter is the global grid quantity $[{\rm M/H}]$, which was compared with homogenised optical literature $[{\rm Fe/H}]$ values in order to validate the near-infrared metallicity scale.

For the primary 14-star homogeneous SpeX sample, the inferred
metallicities reproduce the homogenised optical reference scale with a
mean offset of $-0.054$ dex, a standard deviation of $0.085$ dex, and
a robust scatter of $0.119$ dex. The median posterior uncertainty in
$[{\rm M/H}]$ is approximately $0.22$ dex and the RMS normalised
residual is $0.383$. Including the two additional stars taken from the
IRTF Spectral Library changes these values only slightly, yielding a mean offset of $-0.061$ dex, a scatter of $0.082$ dex, a robust scatter of $0.104$ dex, and an RMS normalised residual of $0.382$, demonstrating the robustness of the inferred metallicity scale. These results show that the $Y+J$ spectra contain useful metallicity information and that the resulting near-infrared 
metallicity scale reproduces the homogenised optical reference scale with a scatter of approximately 0.1 dex.  Within the limits of the present sample, the residuals show no obvious trends with effective temperature, pulsation period, or flux-weighted gravity.

A key outcome of the analysis is the importance of residual modelling. At this spectral resolution, the metallicity information is distributed across blended atomic and molecular features, while the model-data residuals exhibit both short-range correlations and localised problematic wavelength regions. The adopted production likelihood, which combines a regularised short-range covariance component with a local flagged-pixel tolerance, provides stable metallicities and conservative uncertainties while preserving the metallicity-sensitive spectral information.

This validation remains limited by the modest sample size, the use of single-epoch spectra, and the restriction to 
selected $Y+J$ wavelength windows. Future work should therefore expand the calibration sample, test phase dependence, and explore additional near-infrared wavelength regions.

The main long-term application is to Local Group and nearby-galaxy Cepheids, where low-resolution near-infrared spectroscopy could provide a practical route to direct Cepheid metallicities in systems for which large high-resolution optical samples remain difficult to obtain. In this context, the method is complementary to both high-resolution optical Cepheid spectroscopy and abundance studies of blue supergiants. 
The present validation provides a practical foundation for extending direct Cepheid metallicity measurements to larger
and more distant samples and for establishing a homogeneous near-infrared Cepheid metallicity scale for studies of nearby galaxies.

\begin{acknowledgements}
This work is based on observations obtained at the NASA Infrared Telescope Facility, which is operated by the University of Hawai{\textokina}i under contract 80HQTR24DA010 with the National Aeronautics and Space Administration.

The authors wish to recognise and acknowledge the very significant cultural role and reverence that the summit of Maunakea has always had within the Native Hawaiian community. We are most fortunate to have the opportunity to conduct observations from this mountain.

RPK acknowledges support by the Munich Excellence Cluster Origins and the Munich Institute for Astro-, Particle and Biophysics (MIAPbP) both funded by the Deutsche Forschungsgemeinschaft (DFG, German Research Foundation) under the German Excellence Strategy EXC-2094 390783311.
\end{acknowledgements}

\bibliographystyle{aa} 
\bibliography{biblio}  

@ARTICLE{Alvarez1998,
       author = {{Alvarez}, R. and {Plez}, B.},
        title = "{Near-infrared narrow-band photometry of M-giant and Mira stars: models meet observations}",
      journal = {\aap},
         year = 1998,
        month = feb,
       volume = {330},
        pages = {1109-1119},
          doi = {10.48550/arXiv.astro-ph/9710157},
archivePrefix = {arXiv},
       eprint = {astro-ph/9710157},
 primaryClass = {astro-ph},
       adsurl = {https://ui.adsabs.harvard.edu/abs/1998A&A...330.1109A}
}

@ARTICLE{Anderson2016,
       author = {{Anderson}, R.~I. and {Saio}, H. and {Ekstr{\"o}m}, S. and {Georgy}, C. and {Meynet}, G.},
        title = "{On the effect of rotation on populations of classical Cepheids. II. Pulsation analysis for metallicities 0.014, 0.006, and 0.002}",
      journal = {\aap},
         year = 2016,
        month = jun,
       volume = {591},
          eid = {A8},
        pages = {A8},
          doi = {10.1051/0004-6361/201528031},
archivePrefix = {arXiv},
       eprint = {1604.05691},
 primaryClass = {astro-ph.SR},
       adsurl = {https://ui.adsabs.harvard.edu/abs/2016A&A...591A...8A}
}

@ARTICLE{Andrievsky2002,
       author = {{Andrievsky}, S.~M. and {Kovtyukh}, V.~V. and {Luck}, R.~E. and {L{\'e}pine}, J.~R.~D. and {Bersier}, D. and {Maciel}, W.~J. and {Barbuy}, B. and {Klochkova}, V.~G. and {Panchuk}, V.~E. and {Karpischek}, R.~U.},
        title = "{Using Cepheids to determine the galactic abundance gradient. I. The solar neighbourhood}",
      journal = {\aap},
         year = 2002,
        month = jan,
       volume = {381},
        pages = {32-50},
          doi = {10.1051/0004-6361:20011488},
archivePrefix = {arXiv},
       eprint = {astro-ph/0112525},
 primaryClass = {astro-ph},
       adsurl = {https://ui.adsabs.harvard.edu/abs/2002A&A...381...32A}
}

@ARTICLE{Andrievsky2013,
       author = {{Andrievsky}, S.~M. and {L{\'e}pine}, J.~R.~D. and {Korotin}, S.~A. and {Luck}, R.~E. and {Kovtyukh}, V.~V. and {Maciel}, W.~J.},
        title = "{Barium abundances in Cepheids}",
      journal = {\mnras},
         year = 2013,
        month = feb,
       volume = {428},
       number = {4},
        pages = {3252-3261},
          doi = {10.1093/mnras/sts270},
archivePrefix = {arXiv},
       eprint = {1210.6211},
 primaryClass = {astro-ph.GA},
       adsurl = {https://ui.adsabs.harvard.edu/abs/2013MNRAS.428.3252A}
}

@ARTICLE{Bhardwaj2023,
  author  = {{Bhardwaj}, A. and {Riess}, A.~G. and {Catanzaro}, G. and {Trentin}, E. and {Ripepi}, V. and {Rejkuba}, M. and {Marconi}, M. and {Ngeow}, C.-C. and {Macri}, L.~M. and {Romaniello}, M. and {Molinaro}, R. and {Singh}, H.~P. and {Kanbur}, S.~M.},
  title   = {{High-resolution Spectroscopic Metallicities of Milky Way Cepheid Standards and Their Impact on the Leavitt Law and the Hubble Constant}},
  journal = {ApJL},
  year    = {2023},
  volume  = {955},
  number  = {1},
  eid     = {L13},
  pages   = {L13},
  doi     = {10.3847/2041-8213/acf710}
}

@ARTICLE{Berger2018,
       author = {{Berger}, Travis A. and {Kudritzki}, Rolf-Peter and {Urbaneja}, Miguel A. and {Bresolin}, Fabio and {Gieren}, Wolfgang and {Pietrzy{\'n}ski}, Grzegorz and {Przybilla}, Norbert},
        title = "{Quantitative Spectroscopy of Supergiants in the Local Group Dwarf Galaxy IC 1613: Metallicity and Distance}",
      journal = {\apj},
         year = 2018,
        month = jun,
       volume = {860},
       number = {2},
          eid = {130},
        pages = {130},
          doi = {10.3847/1538-4357/aac493},
archivePrefix = {arXiv},
       eprint = {1805.07352},
 primaryClass = {astro-ph.GA},
       adsurl = {https://ui.adsabs.harvard.edu/abs/2018ApJ...860..130B}
}

@ARTICLE{Bersier1996,
       author = {{Bersier}, D. and {Burki}, G.},
        title = "{Fundamental parameters of Cepheids. III. Turbulence variations.}",
      journal = {\aap},
         year = 1996,
        month = feb,
       volume = {306},
        pages = {417},
       adsurl = {https://ui.adsabs.harvard.edu/abs/1996A&A...306..417B}
}

@ARTICLE{Borra2017,
       author = {{Borra}, E.~F. and {Deschatelets}, D.},
        title = "{Measurements of microturbulence of Cepheids using the autocorrelation function}",
      journal = {\mnras},
         year = 2017,
        month = oct,
       volume = {470},
       number = {4},
        pages = {4732-4738},
          doi = {10.1093/mnras/stx1546},
archivePrefix = {arXiv},
       eprint = {1707.00738},
 primaryClass = {astro-ph.SR},
       adsurl = {https://ui.adsabs.harvard.edu/abs/2017MNRAS.470.4732B}
}

@ARTICLE{Bresolin2025,
       author = {{Bresolin}, Fabio and {Kudritzki}, Rolf-Peter and {Urbaneja}, Miguel A. and {Sextl}, Eva and {Riess}, Adam G.},
        title = "{Blue Supergiants in the Pinwheel Galaxy M101: Comparison with H II Region Chemical Abundances, Spectroscopic Distance, and an Independent Determination of the Hubble Constant}",
      journal = {\apj},
         year = 2025,
        month = oct,
       volume = {991},
       number = {2},
          eid = {151},
        pages = {151},
          doi = {10.3847/1538-4357/adfc4c},
archivePrefix = {arXiv},
       eprint = {2508.11837},
 primaryClass = {astro-ph.GA},
       adsurl = {https://ui.adsabs.harvard.edu/abs/2025ApJ...991..151B}
}

@ARTICLE{Breuval2022,
       author = {{Breuval}, Louise and {Riess}, Adam G. and {Kervella}, Pierre and {Anderson}, Richard I. and {Romaniello}, Martino},
        title = "{An Improved Calibration of the Wavelength Dependence of Metallicity on the Cepheid Leavitt Law}",
      journal = {\apj},
         year = 2022,
        month = nov,
       volume = {939},
       number = {2},
          eid = {89},
        pages = {89},
          doi = {10.3847/1538-4357/ac97e2},
archivePrefix = {arXiv},
       eprint = {2205.06280},
 primaryClass = {astro-ph.GA},
       adsurl = {https://ui.adsabs.harvard.edu/abs/2022ApJ...939...89B}
}

@ARTICLE{Casey2016,
       author = {{Casey}, Andrew R.},
        title = "{sick: The Spectroscopic Inference Crank}",
      journal = {\apjs},
         year = 2016,
        month = mar,
       volume = {223},
       number = {1},
          eid = {8},
        pages = {8},
          doi = {10.3847/0067-0049/223/1/8},
archivePrefix = {arXiv},
       eprint = {1603.03043},
 primaryClass = {astro-ph.IM},
       adsurl = {https://ui.adsabs.harvard.edu/abs/2016ApJS..223....8C}
}

@ARTICLE{Catanzaro2026,
       author = {{Catanzaro}, G. and {Bhardwaj}, A. and {Ripepi}, V. and {Trentin}, E. and {Marconi}, M. and {Romaniello}, M. and {Matsunaga}, N. and {De Somma}, G. and {Sicignano}, T. and {Musella}, I. and {Luongo}, E. and {Testa}, V. and {Soung-Chul}, Y.},
        title = "{Cepheid Metallicity in the Leavitt Law (C─MetaLL) survey: VII. High-resolution IGRINS spectroscopy of 23 classical Cepheids: Validating NIR abundances}",
      journal = {\aap},
         year = 2026,
        month = feb,
       volume = {706},
          eid = {A225},
        pages = {A225},
          doi = {10.1051/0004-6361/202558020},
archivePrefix = {arXiv},
       eprint = {2511.06264},
 primaryClass = {astro-ph.SR},
       adsurl = {https://ui.adsabs.harvard.edu/abs/2026A&A...706A.225C}
}

@ARTICLE{Cirasuolo2020,
       author = {{Cirasuolo}, M. and {Fairley}, A. and {Rees}, P. and {Gonzalez}, O.~A. and {Taylor}, W. and {Maiolino}, R. and {Afonso}, J. and {Evans}, C. and {Flores}, H. and {Lilly}, S. and {Oliva}, E. and {Paltani}, S. and {Vanzi}, L. and {Abreu}, M. and {Accardo}, M. and {Adams}, N. and {{\'A}lvarez M{\'e}ndez}, D. and {Amans}, J.-P. and {Amarantidis}, S. and {Atek}, H. and {Atkinson}, D. and {Banerji}, M. and {Barrett}, J. and {Barrientos}, F. and {Bauer}, F. and {Beard}, S. and {B{\'e}chet}, C. and {Belfiore}, A. and {Bellazzini}, M. and {Benoist}, C. and {Best}, P. and {Biazzo}, K. and {Black}, M. and {Boettger}, D. and {Bonifacio}, P. and {Bowler}, R. and {Bragaglia}, A. and {Brierley}, S. and {Brinchmann}, J. and {Brinkmann}, M. and {Buat}, V. and {Buitrago}, F. and {Burgarella}, D. and {Burningham}, B. and {Buscher}, D. and {Cabral}, A. and {Caffau}, E. and {Cardoso}, L. and {Carnall}, A. and {Carollo}, M. and {Castillo}, R. and {Castignani}, G. and {Catelan}, M. and {Cicone}, C. and {Cimatti}, A. and {Cioni}, M.-R.~L. and {Clementini}, G. and {Cochrane}, W. and {Coelho}, J. and {Colling}, M. and {Contini}, T. and {Contreras}, R. and {Conzelmann}, R. and {Cresci}, G. and {Cropper}, M. and {Cucciati}, O. and {Cullen}, F. and {Cumani}, C. and {Curti}, M. and {Da Silva}, A. and {Daddi}, E. and {Dalessandro}, E. and {Dalessio}, F. and {Dauvin}, L. and {Davidson}, G. and {de Laverny}, P. and {Delplancke-Str{\"o}bele}, F. and {De Lucia}, G. and {Del Vecchio}, C. and {Dessauges-Zavadsky}, M. and {Di Matteo}, P. and {Dole}, H. and {Drass}, H. and {Dunlop}, J. and {D{\"u}nner}, R. and {Eales}, S. and {Ellis}, R. and {Enriques}, B. and {Fasola}, G. and {Ferguson}, A. and {Ferruzzi}, D. and {Fisher}, M. and {Flores}, M. and {Fontana}, A. and {Forchi}, V. and {Francois}, P. and {Franzetti}, P. and {Gargiulo}, A. and {Garilli}, B. and {Gaudemard}, J. and {Gieles}, M. and {Gilmore}, G. and {Ginolfi}, M. and {Gomes}, J.~M. and {Guinouard}, I. and {Gutierrez}, P. and {Haigron}, R. and {Hammer}, F. and {Hammersley}, P. and {Haniff}, C. and {Harrison}, C. and {Haywood}, M. and {Hill}, V. and {Hubin}, N. and {Humphrey}, A. and {Ibata}, R. and {Infante}, L. and {Ives}, D. and {Ivison}, R. and {Iwert}, O. and {Jablonka}, P. and {Jakob}, G. and {Jarvis}, M. and {King}, D. and {Kneib}, J.-P. and {Laporte}, P. and {Lawrence}, A. and {Lee}, D. and {Li Causi}, G. and {Lorenzoni}, S. and {Lucatello}, S. and {Luco}, Y. and {Macleod}, A. and {Magliocchetti}, M. and {Magrini}, L. and {Mainieri}, V. and {Maire}, C. and {Mannucci}, F. and {Martin}, N. and {Matute}, I. and {Maurogordato}, S. and {McGee}, S. and {Mcleod}, D. and {McLure}, R. and {McMahon}, R. and {Melse}, B.-T. and {Messias}, H. and {Mucciarelli}, A. and {Nisini}, B. and {Nix}, J. and {Norberg}, P. and {Oesch}, P. and {Oliveira}, A. and {Origlia}, L. and {Padilla}, N. and {Palsa}, R. and {Pancino}, E. and {Papaderos}, P. and {Pappalardo}, C. and {Parry}, I. and {Pasquini}, L. and {Peacock}, J. and {Pedichini}, F. and {Pello}, R. and {Peng}, Y. and {Pentericci}, L. and {Pfuhl}, O. and {Piazzesi}, R. and {Popovic}, D. and {Pozzetti}, L. and {Puech}, M. and {Puzia}, T. and {Raichoor}, A. and {Randich}, S. and {Recio-Blanco}, A. and {Reis}, S. and {Reix}, F. and {Renzini}, A. and {Rodrigues}, M. and {Rojas}, F. and {Rojas-Arriagada}, {\'A}. and {Rota}, S. and {Royer}, F. and {Sacco}, G. and {Sanchez-Janssen}, R. and {Sanna}, N. and {Santos}, P. and {Sarzi}, M. and {Schaerer}, D. and {Schiavon}, R. and {Schnell}, R. and {Schultheis}, M. and {Scodeggio}, M. and {Serjeant}, S. and {Shen}, T.-C. and {Simmonds}, C. and {Smoker}, J. and {Sobral}, D. and {Sordet}, M. and {Sp{\'e}rone}, D.},
        title = "{MOONS: The New Multi-Object Spectrograph for the VLT}",
      journal = {The Messenger},
         year = 2020,
        month = jun,
       volume = {180},
        pages = {10-17},
          doi = {10.18727/0722-6691/5195},
archivePrefix = {arXiv},
       eprint = {2009.00628},
 primaryClass = {astro-ph.IM},
       adsurl = {https://ui.adsabs.harvard.edu/abs/2020Msngr.180...10C}
}

@ARTICLE{Cushing2004,
       author = {{Cushing}, Michael C. and {Vacca}, William D. and {Rayner}, John T.},
        title = "{Spextool: A Spectral Extraction Package for SpeX, a 0.8-5.5 Micron Cross-Dispersed Spectrograph}",
      journal = {\pasp},
         year = 2004,
        month = apr,
       volume = {116},
       number = {818},
        pages = {362-376},
          doi = {10.1086/382907},
       adsurl = {https://ui.adsabs.harvard.edu/abs/2004PASP..116..362C}
}

@ARTICLE{Czekala2015,
       author = {{Czekala}, Ian and {Andrews}, Sean M. and {Mandel}, Kaisey S. and {Hogg}, David W. and {Green}, Gregory M.},
        title = "{Constructing a Flexible Likelihood Function for Spectroscopic Inference}",
      journal = {\apj},
         year = 2015,
        month = oct,
       volume = {812},
       number = {2},
          eid = {128},
        pages = {128},
          doi = {10.1088/0004-637X/812/2/128},
archivePrefix = {arXiv},
       eprint = {1412.5177},
 primaryClass = {astro-ph.SR},
       adsurl = {https://ui.adsabs.harvard.edu/abs/2015ApJ...812..128C}
}

@ARTICLE{Davies2010,
       author = {{Davies}, Ben and {Kudritzki}, Rolf-Peter and {Figer}, Donald F.},
        title = "{The potential of red supergiants as extragalactic abundance probes at low spectral resolution}",
      journal = {\mnras},
         year = 2010,
        month = sep,
       volume = {407},
       number = {2},
        pages = {1203-1211},
          doi = {10.1111/j.1365-2966.2010.16965.x},
archivePrefix = {arXiv},
       eprint = {1005.1008},
 primaryClass = {astro-ph.CO},
       adsurl = {https://ui.adsabs.harvard.edu/abs/2010MNRAS.407.1203D}
}

@ARTICLE{Freedman2010,
       author = {{Freedman}, Wendy L. and {Madore}, Barry F.},
        title = "{The Hubble Constant}",
      journal = {\araa},
         year = 2010,
        month = sep,
       volume = {48},
        pages = {673-710},
          doi = {10.1146/annurev-astro-082708-101829},
archivePrefix = {arXiv},
       eprint = {1004.1856},
 primaryClass = {astro-ph.CO},
       adsurl = {https://ui.adsabs.harvard.edu/abs/2010ARA&A..48..673F}
}

@ARTICLE{Gazak2014,
       author = {{Gazak}, J. Zachary and {Davies}, Ben and {Kudritzki}, Rolf and {Bergemann}, Maria and {Plez}, Bertrand},
        title = "{Quantitative Spectroscopic J-band study of Red Supergiants in Perseus OB-1}",
      journal = {\apj},
         year = 2014,
        month = jun,
       volume = {788},
       number = {1},
          eid = {58},
        pages = {58},
          doi = {10.1088/0004-637X/788/1/58},
archivePrefix = {arXiv},
       eprint = {1404.5713},
 primaryClass = {astro-ph.SR},
       adsurl = {https://ui.adsabs.harvard.edu/abs/2014ApJ...788...58G}
}

@ARTICLE{Geballe2011,
       author = {{Geballe}, T.~R. and {Najarro}, F. and {Figer}, D.~F. and {Schlegelmilch}, B.~W. and {de La Fuente}, D.},
        title = "{Infrared diffuse interstellar bands in the Galactic Centre region}",
      journal = {\nat},
         year = 2011,
        month = nov,
       volume = {479},
       number = {7372},
        pages = {200-202},
          doi = {10.1038/nature10527},
archivePrefix = {arXiv},
       eprint = {1111.0613},
 primaryClass = {astro-ph.GA},
       adsurl = {https://ui.adsabs.harvard.edu/abs/2011Natur.479..200G}
}

@ARTICLE{Genovali2014,
       author = {{Genovali}, K. and {Lemasle}, B. and {Bono}, G. and {Romaniello}, M. and {Fabrizio}, M. and {Ferraro}, I. and {Iannicola}, G. and {Laney}, C.~D. and {Nonino}, M. and {Bergemann}, M. and {Buonanno}, R. and {Fran{\c{c}}ois}, P. and {Inno}, L. and {Kudritzki}, R.-P. and {Matsunaga}, N. and {Pedicelli}, S. and {Primas}, F. and {Th{\'e}venin}, F.},
        title = "{On the fine structure of the Cepheid metallicity gradient in the Galactic thin disk}",
      journal = {\aap},
         year = 2014,
        month = jun,
       volume = {566},
          eid = {A37},
        pages = {A37},
          doi = {10.1051/0004-6361/201323198},
archivePrefix = {arXiv},
       eprint = {1403.6128},
 primaryClass = {astro-ph.GA},
       adsurl = {https://ui.adsabs.harvard.edu/abs/2014A&A...566A..37G}
}

@ARTICLE{Groenewegen2020,
       author = {{Groenewegen}, M.~A.~T.},
        title = "{The flux-weighted gravity-luminosity relation of Galactic classical Cepheids}",
      journal = {\aap},
         year = 2020,
        month = aug,
       volume = {640},
          eid = {A113},
        pages = {A113},
          doi = {10.1051/0004-6361/202038292},
archivePrefix = {arXiv},
       eprint = {2007.02148},
 primaryClass = {astro-ph.SR},
       adsurl = {https://ui.adsabs.harvard.edu/abs/2020A&A...640A.113G}
}

@ARTICLE{Groenewegen2023,
       author = {{Groenewegen}, M.~A.~T. and {Lub}, J.},
        title = "{Spectral energy distributions of classical Cepheids in the Magellanic Clouds}",
      journal = {\aap},
         year = 2023,
        month = aug,
       volume = {676},
          eid = {A136},
        pages = {A136},
          doi = {10.1051/0004-6361/202346062},
archivePrefix = {arXiv},
       eprint = {2307.07559},
 primaryClass = {astro-ph.SR},
       adsurl = {https://ui.adsabs.harvard.edu/abs/2023A&A...676A.136G}
}

@ARTICLE{Gustafsson2008,
       author = {{Gustafsson}, B. and {Edvardsson}, B. and {Eriksson}, K. and {J{\o}rgensen}, U.~G. and {Nordlund}, {\r{A}}. and {Plez}, B.},
        title = "{A grid of MARCS model atmospheres for late-type stars. I. Methods and general properties}",
      journal = {\aap},
         year = 2008,
        month = aug,
       volume = {486},
       number = {3},
        pages = {951-970},
          doi = {10.1051/0004-6361:200809724},
archivePrefix = {arXiv},
       eprint = {0805.0554},
 primaryClass = {astro-ph},
       adsurl = {https://ui.adsabs.harvard.edu/abs/2008A&A...486..951G}
}

@ARTICLE{Hosek2014,
       author = {{Hosek}, Jr., Matthew W. and {Kudritzki}, Rolf-Peter and {Bresolin}, Fabio and {Urbaneja}, Miguel A. and {Evans}, Christopher J. and {Pietrzy{\'n}ski}, Grzegorz and {Gieren}, Wolfgang and {Przybilla}, Norbert and {Carraro}, Giovanni},
        title = "{Quantitative Spectroscopy of Blue Supergiants in Metal-poor Dwarf Galaxy NGC 3109}",
      journal = {\apj},
         year = 2014,
        month = apr,
       volume = {785},
       number = {2},
          eid = {151},
        pages = {151},
          doi = {10.1088/0004-637X/785/2/151},
archivePrefix = {arXiv},
       eprint = {1402.6358},
 primaryClass = {astro-ph.GA},
       adsurl = {https://ui.adsabs.harvard.edu/abs/2014ApJ...785..151H}
}

@ARTICLE{Inno2019,
       author = {{Inno}, L. and {Urbaneja}, M.~A. and {Matsunaga}, N. and {Bono}, G. and {Nonino}, M. and {Debattista}, V.~P. and {Sormani}, M.~C. and {Bergemann}, M. and {da Silva}, R. and {Lemasle}, B. and {Romaniello}, M. and {Rix}, H.-W.},
        title = "{First metallicity determination from near-infrared spectra for five obscured Cepheids discovered in the inner disc}",
      journal = {\mnras},
         year = 2019,
        month = jan,
       volume = {482},
       number = {1},
        pages = {83-97},
          doi = {10.1093/mnras/sty2661},
archivePrefix = {arXiv},
       eprint = {1805.03212},
 primaryClass = {astro-ph.GA},
       adsurl = {https://ui.adsabs.harvard.edu/abs/2019MNRAS.482...83I}
}

@book{Jolliffe2002,
  author    = {Jolliffe, I. T.},
  title     = {Principal Component Analysis},
  edition   = {2},
  publisher = {Springer},
  address   = {New York},
  year      = {2002}
}

@ARTICLE{Kovtyukh2016,
       author = {{Kovtyukh}, V. and {Lemasle}, B. and {Chekhonadskikh}, F. and {Bono}, G. and {Matsunaga}, N. and {Yushchenko}, A. and {Anderson}, R.~I. and {Belik}, S. and {da Silva}, R. and {Inno}, L.},
        title = "{The chemical composition of Galactic beat Cepheids}",
      journal = {\mnras},
         year = 2016,
        month = aug,
       volume = {460},
       number = {2},
        pages = {2077-2086},
          doi = {10.1093/mnras/stw1113},
       adsurl = {https://ui.adsabs.harvard.edu/abs/2016MNRAS.460.2077K}
}

@ARTICLE{Kudritzki2003,
       author = {{Kudritzki}, Rolf P. and {Bresolin}, Fabio and {Przybilla}, Norbert},
        title = "{A New Extragalactic Distance Determination Method Using the Flux-weighted Gravity of Late B and Early A Supergiants}",
      journal = {\apjl},
         year = 2003,
        month = jan,
       volume = {582},
       number = {2},
        pages = {L83-L86},
          doi = {10.1086/367690},
archivePrefix = {arXiv},
       eprint = {astro-ph/0212042},
 primaryClass = {astro-ph},
       adsurl = {https://ui.adsabs.harvard.edu/abs/2003ApJ...582L..83K}
}

@ARTICLE{Kudritzki2008,
       author = {{Kudritzki}, R.~P. and {Urbaneja}, M.~A. and {Bresolin}, F. and {Przybilla}, N.},
        title = "{Extragalactic stellar astronomy with the brightest stars in the universe}",
      journal = {Physica Scripta Volume T},
         year = 2008,
        month = dec,
       volume = {133},
          eid = {014039},
        pages = {014039},
          doi = {10.1088/0031-8949/2008/T133/014039},
archivePrefix = {arXiv},
       eprint = {0803.3656},
 primaryClass = {astro-ph},
       adsurl = {https://ui.adsabs.harvard.edu/abs/2008PhST..133a4039K}
}

@ARTICLE{Kudritzki2024,
       author = {{Kudritzki}, Rolf-Peter and {Urbaneja}, Miguel A. and {Bresolin}, Fabio and {Macri}, Lucas M. and {Yuan}, Wenlong and {Li}, Siyang and {Anand}, Gagandeep S. and {Riess}, Adam G.},
        title = "{The Hubble Constant Anchor Galaxy NGC 4258: Metallicity and Distance from Blue Supergiants}",
      journal = {\apj},
         year = 2024,
        month = dec,
       volume = {977},
       number = {2},
          eid = {217},
        pages = {217},
          doi = {10.3847/1538-4357/ad9279},
archivePrefix = {arXiv},
       eprint = {2411.07974},
 primaryClass = {astro-ph.GA},
       adsurl = {https://ui.adsabs.harvard.edu/abs/2024ApJ...977..217K}
}

@ARTICLE{Kupka1999,
  author  = {{Kupka}, F. and {Piskunov}, N.~E. and {Ryabchikova}, T.~A. and {Stempels}, H.~C. and {Weiss}, W.~W.},
  title   = {{VALD-2: Progress of the Vienna Atomic Line Data Base}},
  journal = {Astronomy and Astrophysics Supplement Series},
  year    = {1999},
  volume  = {138},
  pages   = {119--133},
  doi     = {10.1051/aas:1999267}
}

@article{Lee1999,
  author  = {Lee, D. D. and Seung, H. S.},
  title   = {Learning the Parts of Objects by Non-negative Matrix Factorization},
  journal = {Nature},
  volume  = {401},
  pages   = {788--791},
  year    = {1999},
  doi     = {10.1038/44565}
}

@article{Lee2001,
  author  = {Lee, D. D. and Seung, H. S.},
  title   = {Algorithms for Non-negative Matrix Factorization},
  journal = {Advances in Neural Information Processing Systems},
  volume  = {13},
  pages   = {556--562},
  year    = {2001}
}

@ARTICLE{Lemasle2017,
       author = {{Lemasle}, B. and {Groenewegen}, M.~A.~T. and {Grebel}, E.~K. and {Bono}, G. and {Fiorentino}, G. and {Fran{\c{c}}ois}, P. and {Inno}, L. and {Kovtyukh}, V.~V. and {Matsunaga}, N. and {Pedicelli}, S. and {Primas}, F. and {Pritchard}, J. and {Romaniello}, M. and {da Silva}, R.},
        title = "{Detailed chemical composition of classical Cepheids in the LMC cluster NGC 1866 and in the field of the SMC}",
      journal = {\aap},
         year = 2017,
        month = dec,
       volume = {608},
          eid = {A85},
        pages = {A85},
          doi = {10.1051/0004-6361/201731370},
archivePrefix = {arXiv},
       eprint = {1709.03083},
 primaryClass = {astro-ph.GA},
       adsurl = {https://ui.adsabs.harvard.edu/abs/2017A&A...608A..85L}
}

@ARTICLE{Luck2011,
       author = {{Luck}, R. Earle and {Lambert}, David L.},
        title = "{The Distribution of the Elements in the Galactic Disk. III. A Reconsideration of Cepheids from l = 30{\textdegree} to 250{\textdegree}}",
      journal = {\aj},
         year = 2011,
        month = oct,
       volume = {142},
       number = {4},
          eid = {136},
        pages = {136},
          doi = {10.1088/0004-6256/142/4/136},
archivePrefix = {arXiv},
       eprint = {1108.1947},
 primaryClass = {astro-ph.GA},
       adsurl = {https://ui.adsabs.harvard.edu/abs/2011AJ....142..136L}
}

@ARTICLE{Luck2018,
       author = {{Luck}, R. Earle},
        title = "{Cepheid Abundances: Multiphase Results and Spatial Gradients}",
      journal = {\aj},
         year = 2018,
        month = oct,
       volume = {156},
       number = {4},
          eid = {171},
        pages = {171},
          doi = {10.3847/1538-3881/aadcac},
archivePrefix = {arXiv},
       eprint = {1808.05863},
 primaryClass = {astro-ph.SR},
       adsurl = {https://ui.adsabs.harvard.edu/abs/2018AJ....156..171L}
}

@ARTICLE{Matsunaga2023,
       author = {{Matsunaga}, Noriyuki and {Taniguchi}, Daisuke and {Elgueta}, Scarlet S. and {Tsujimoto}, Takuji and {Baba}, Junichi and {McWilliam}, Andrew and {Otsubo}, Shogo and {Sarugaku}, Yuki and {Takeuchi}, Tomomi and {Katoh}, Haruki and {Hamano}, Satoshi and {Ikeda}, Yuji and {Kawakita}, Hideyo and {Hull}, Charlie and {Albarrac{\'\i}n}, Rogelio and {Bono}, Giuseppe and {D'Orazi}, Valentina},
        title = "{Metallicities of Classical Cepheids in the Inner Galactic Disk}",
      journal = {\apj},
         year = 2023,
        month = sep,
       volume = {954},
       number = {2},
          eid = {198},
        pages = {198},
          doi = {10.3847/1538-4357/aced93},
archivePrefix = {arXiv},
       eprint = {2308.02853},
 primaryClass = {astro-ph.SR},
       adsurl = {https://ui.adsabs.harvard.edu/abs/2023ApJ...954..198M}
}

@ARTICLE{Nunnari2026,
       author = {{Nunnari}, A. and {D'Orazi}, V. and {Fiorentino}, G. and {Braga}, V.~F. and {Bono}, G. and {Fabrizio}, M. and {J{\"o}nsson}, H. and {Kudritzki}, R.-P. and {da Silva}, R. and {Bergemann}, M. and {Poggio}, E. and {Otto}, J.~M. and {Baeza-Villagra}, K. and {Bragaglia}, A. and {Ceci}, G. and {Dall'Ora}, M. and {Inno}, L. and {Lardo}, C. and {Matsunaga}, N. and {Monelli}, M. and {S{\'a}nchez-Benavente}, M. and {Sneden}, C. and {Tantalo}, M. and {Th{\'e}v{\'e}nin}, F. and {Kovtyukh}, V. and {Di Criscienzo}, M. and {B{\"o}cek Topcu}, G.},
        title = "{Classical Cepheids in the Galactic thin disk: I. Abundance gradients via non-local thermodynamic equilibrium spectral analysis}",
      journal = {\aap},
         year = 2026,
        month = mar,
       volume = {708},
          eid = {A17},
        pages = {A17},
          doi = {10.1051/0004-6361/202558288},
archivePrefix = {arXiv},
       eprint = {2511.22491},
 primaryClass = {astro-ph.GA},
       adsurl = {https://ui.adsabs.harvard.edu/abs/2026A&A...708A..17N}
}

@software{Plez2012,
       author = {{Plez}, B.},
        title = "{Turbospectrum: Code for spectral synthesis}",
 howpublished = {Astrophysics Source Code Library, record ascl:1205.004},
         year = 2012,
        month = may,
          eid = {ascl:1205.004},
archivePrefix = {ascl},
       eprint = {1205.004},
       adsurl = {https://ui.adsabs.harvard.edu/abs/2012ascl.soft05004P}
}

@ARTICLE{Rayner2009,
       author = {{Rayner}, John T. and {Cushing}, Michael C. and {Vacca}, William D.},
        title = "{The Infrared Telescope Facility (IRTF) Spectral Library: Cool Stars}",
      journal = {\apjs},
         year = 2009,
        month = dec,
       volume = {185},
       number = {2},
        pages = {289-432},
          doi = {10.1088/0067-0049/185/2/289},
archivePrefix = {arXiv},
       eprint = {0909.0818},
 primaryClass = {astro-ph.SR},
       adsurl = {https://ui.adsabs.harvard.edu/abs/2009ApJS..185..289R}
}

@ARTICLE{Riess2022,
       author = {{Riess}, Adam G. and {Yuan}, Wenlong and {Macri}, Lucas M. and {Scolnic}, Dan and {Brout}, Dillon and {Casertano}, Stefano and {Jones}, David O. and {Murakami}, Yukei and {Anand}, Gagandeep S. and {Breuval}, Louise and {Brink}, Thomas G. and {Filippenko}, Alexei V. and {Hoffmann}, Samantha and {Jha}, Saurabh W. and {D'arcy Kenworthy}, W. and {Mackenty}, John and {Stahl}, Benjamin E. and {Zheng}, WeiKang},
        title = "{A Comprehensive Measurement of the Local Value of the Hubble Constant with 1 km s$^{-1}$ Mpc$^{-1}$ Uncertainty from the Hubble Space Telescope and the SH0ES Team}",
      journal = {\apjl},
         year = 2022,
        month = jul,
       volume = {934},
       number = {1},
          eid = {L7},
        pages = {L7},
          doi = {10.3847/2041-8213/ac5c5b},
archivePrefix = {arXiv},
       eprint = {2112.04510},
 primaryClass = {astro-ph.CO},
       adsurl = {https://ui.adsabs.harvard.edu/abs/2022ApJ...934L...7R}
}

@ARTICLE{Ripepi2020,
       author = {{Ripepi}, V. and {Catanzaro}, G. and {Molinaro}, R. and {Marconi}, M. and {Clementini}, G. and {Cusano}, F. and {De Somma}, G. and {Leccia}, S. and {Musella}, I. and {Testa}, V.},
        title = "{Period-luminosity-metallicity relation of classical Cepheids}",
      journal = {\aap},
         year = 2020,
        month = oct,
       volume = {642},
          eid = {A230},
        pages = {A230},
          doi = {10.1051/0004-6361/202038714},
archivePrefix = {arXiv},
       eprint = {2008.04608},
 primaryClass = {astro-ph.SR},
       adsurl = {https://ui.adsabs.harvard.edu/abs/2020A&A...642A.230R}
}

@ARTICLE{Romaniello2022,
       author = {{Romaniello}, Martino and {Riess}, Adam and {Mancino}, Sara and {Anderson}, Richard I. and {Freudling}, Wolfram and {Kudritzki}, Rolf-Peter and {Macr{\`\i}}, Lucas and {Mucciarelli}, Alessio and {Yuan}, Wenlong},
        title = "{The iron and oxygen content of LMC Classical Cepheids and its implications for the extragalactic distance scale and Hubble constant. Equivalent width analysis with Kurucz stellar atmosphere models}",
      journal = {\aap},
         year = 2022,
        month = feb,
       volume = {658},
          eid = {A29},
        pages = {A29},
          doi = {10.1051/0004-6361/202142441},
archivePrefix = {arXiv},
       eprint = {2110.08860},
 primaryClass = {astro-ph.CO},
       adsurl = {https://ui.adsabs.harvard.edu/abs/2022A&A...658A..29R}
}

@ARTICLE{Ryabchikova2015,
  author  = {{Ryabchikova}, T. and {Piskunov}, N. and {Kurucz}, R.~L. and {Stempels}, H.~C. and {Heiter}, U. and {Pakhomov}, Yu. and {Barklem}, P.~S.},
  title   = {{A major upgrade of the VALD database}},
  journal = {Physica Scripta},
  year    = {2015},
  volume  = {90},
  number  = {5},
  eid     = {054005},
  pages   = {054005},
  doi     = {10.1088/0031-8949/90/5/054005}
}

@ARTICLE{Ryde2009,
       author = {{Ryde}, N. and {Edvardsson}, B. and {Gustafsson}, B. and {Eriksson}, K. and {K{\"a}ufl}, H.~U. and {Siebenmorgen}, R. and {Smette}, A.},
        title = "{Abundances in bulge stars from high-resolution, near-IR spectra. I. The CNO elements observed during the science verification of CRIRES at VLT}",
      journal = {\aap},
         year = 2009,
        month = mar,
       volume = {496},
       number = {3},
        pages = {701-712},
          doi = {10.1051/0004-6361/200811070},
archivePrefix = {arXiv},
       eprint = {0902.2124},
 primaryClass = {astro-ph.SR},
       adsurl = {https://ui.adsabs.harvard.edu/abs/2009A&A...496..701R}
}

@ARTICLE{Ting2019,
       author = {{Ting}, Yuan-Sen and {Conroy}, Charlie and {Rix}, Hans-Walter and {Cargile}, Phillip},
        title = "{The Payne: Self-consistent ab initio Fitting of Stellar Spectra}",
      journal = {\apj},
         year = 2019,
        month = jul,
       volume = {879},
       number = {2},
          eid = {69},
        pages = {69},
          doi = {10.3847/1538-4357/ab2331},
archivePrefix = {arXiv},
       eprint = {1804.01530},
 primaryClass = {astro-ph.SR},
       adsurl = {https://ui.adsabs.harvard.edu/abs/2019ApJ...879...69T}
}

@ARTICLE{Trentin2024,
       author = {{Trentin}, E. and {Catanzaro}, G. and {Ripepi}, V. and {Alonso-Santiago}, J. and {Molinaro}, R. and {Storm}, J. and {De Somma}, G. and {Marconi}, M. and {Bhardwaj}, A. and {Gatto}, M. and {Musella}, I. and {Testa}, V.},
        title = "{Cepheid Metallicity in the Leavitt Law (C-MetaLL) survey: VI. Radial abundance gradients of 29 chemical species in the Milky Way disc}",
      journal = {\aap},
         year = 2024,
        month = oct,
       volume = {690},
          eid = {A246},
        pages = {A246},
          doi = {10.1051/0004-6361/202450376},
archivePrefix = {arXiv},
       eprint = {2404.17299},
 primaryClass = {astro-ph.GA},
       adsurl = {https://ui.adsabs.harvard.edu/abs/2024A&A...690A.246T}
}

@ARTICLE{U2009,
       author = {{U}, Vivian and {Urbaneja}, Miguel A. and {Kudritzki}, Rolf-Peter and {Jacobs}, Bradley A. and {Bresolin}, Fabio and {Przybilla}, Norbert},
        title = "{A New Distance to M33 Using Blue Supergiants and the FGLR Method}",
      journal = {\apj},
         year = 2009,
        month = oct,
       volume = {704},
       number = {2},
        pages = {1120-1134},
          doi = {10.1088/0004-637X/704/2/1120},
archivePrefix = {arXiv},
       eprint = {0909.0032},
 primaryClass = {astro-ph.CO},
       adsurl = {https://ui.adsabs.harvard.edu/abs/2009ApJ...704.1120U}
}

@ARTICLE{Urbaneja2008,
       author = {{Urbaneja}, Miguel A. and {Kudritzki}, Rolf-Peter and {Bresolin}, Fabio and {Przybilla}, Norbert and {Gieren}, Wolfgang and {Pietrzy{\'n}ski}, Grzegorz},
        title = "{The Araucaria Project: The Local Group Galaxy WLM{\textemdash}Distance and Metallicity from Quantitative Spectroscopy of Blue Supergiants}",
      journal = {\apj},
         year = 2008,
        month = sep,
       volume = {684},
       number = {1},
        pages = {118-135},
          doi = {10.1086/590334},
archivePrefix = {arXiv},
       eprint = {0805.3555},
 primaryClass = {astro-ph},
       adsurl = {https://ui.adsabs.harvard.edu/abs/2008ApJ...684..118U}
}

@ARTICLE{Urbaneja2017,
       author = {{Urbaneja}, M.~A. and {Kudritzki}, R.-P. and {Gieren}, W. and {Pietrzy{\'n}ski}, G. and {Bresolin}, F. and {Przybilla}, N.},
        title = "{LMC Blue Supergiant Stars and the Calibration of the Flux-weighted Gravity-Luminosity Relationship}",
      journal = {\aj},
         year = 2017,
        month = sep,
       volume = {154},
       number = {3},
          eid = {102},
        pages = {102},
          doi = {10.3847/1538-3881/aa79a8},
archivePrefix = {arXiv},
       eprint = {1706.03967},
 primaryClass = {astro-ph.SR},
       adsurl = {https://ui.adsabs.harvard.edu/abs/2017AJ....154..102U}
}

@ARTICLE{Urbaneja2026,
       author = {{Urbaneja}, M.~A.},
        title = "{A statistical framework for the quantitative spectroscopy of luminous blue stars}",
      journal = {\aap},
         year = 2026,
        month = mar,
       volume = {707},
          eid = {A249},
        pages = {A249},
          doi = {10.1051/0004-6361/202558001},
archivePrefix = {arXiv},
       eprint = {2601.01491},
 primaryClass = {astro-ph.SR},
       adsurl = {https://ui.adsabs.harvard.edu/abs/2026A&A...707A.249U}
}

@ARTICLE{Vacca2003,
       author = {{Vacca}, William D. and {Cushing}, Michael C. and {Rayner}, John T.},
        title = "{A Method of Correcting Near-Infrared Spectra for Telluric Absorption}",
      journal = {\pasp},
         year = 2003,
        month = mar,
       volume = {115},
       number = {805},
        pages = {389-409},
          doi = {10.1086/346193},
archivePrefix = {arXiv},
       eprint = {astro-ph/0211255},
 primaryClass = {astro-ph},
       adsurl = {https://ui.adsabs.harvard.edu/abs/2003PASP..115..389V}
}

@ARTICLE{Villaume2017,
       author = {{Villaume}, Alexa and {Conroy}, Charlie and {Johnson}, Benjamin and {Rayner}, John and {Mann}, Andrew W. and {van Dokkum}, Pieter},
        title = "{The Extended IRTF Spectral Library: Expanded Coverage in Metallicity, Temperature, and Surface Gravity}",
      journal = {\apjs},
         year = 2017,
        month = jun,
       volume = {230},
       number = {2},
          eid = {23},
        pages = {23},
          doi = {10.3847/1538-4365/aa72ed},
archivePrefix = {arXiv},
       eprint = {1705.08906},
 primaryClass = {astro-ph.SR},
       adsurl = {https://ui.adsabs.harvard.edu/abs/2017ApJS..230...23V}
}

@ARTICLE{Ebenbichler22,
       author = {{Ebenbichler}, A. and {Postel}, A. and {Przybilla}, N. and {Seifahrt}, A. and {We{\ss}mayer}, D. and {Kausch}, W. and {Firnstein}, M. and {Butler}, K. and {Kaufer}, A. and {Linnartz}, H.},
        title = "{CRIRES high-resolution near-infrared spectroscopy of diffuse interstellar band profiles. Detection of 12 new DIBs in the YJ band and the introduction of a combined ISM sight line and stellar analysis approach}",
      journal = {\aap},
         year = 2022,
        month = jun,
       volume = {662},
          eid = {A81},
        pages = {A81},
          doi = {10.1051/0004-6361/202142990},
archivePrefix = {arXiv},
       eprint = {2203.12562},
 primaryClass = {astro-ph.GA},
       adsurl = {https://ui.adsabs.harvard.edu/abs/2022A&A...662A..81E}
}

@INPROCEEDINGS{Sturm2024,
       author = {{Sturm}, E. and {Davies}, R. and {Alves}, J. and {Cl{\'e}net}, Y. and {Kotilainen}, J. and {Monna}, A. and {Nicklas}, H. and {Pott}, J.-U. and {Tolstoy}, E. and {Vulcani}, B. and {Achren}, J. and {Annadevara}, S. and {Anwand-Heerwart}, H. and {Arcidiacono}, C. and {Barboza}, S. and {Barl}, L. and {Baudoz}, P. and {Bender}, R. and {Bezawada}, N. and {Biondi}, F. and {Bizenberger}, P. and {Blin}, A. and {Bon{\'e}}, A. and {Bonifacio}, P. and {Borgo}, B. and {Born}, J. van den and {Buey}, T. and {Cao}, Y. and {Chapron}, F. and {Chauvin}, G. and {Chemla}, F. and {Cloiseau}, K. and {Cohen}, M. and {Colin}, C. and {Czoske}, O. and {Dette}, J.-O. and {Deysenroth}, M. and {Dijkstra}, E. and {Dreizler}, S. and {Dupuis}, O. and {Egmond}, G. van and {Eisenhauer}, F. and {Elswijk}, E. and {Emslander}, A. and {Fabricius}, M. and {Fasola}, G. and {Ferreira}, F. and {F{\"o}rster Schreiber}, N. and {Fontana}, A. and {Gaudemard}, J. and {Gautherot}, N. and {Gendron}, E. and {Gennet}, C. and {Genzel}, R. and {Ghouchou}, L. and {Gillessen}, S. and {Gratadour}, D. and {Grazian}, A. and {Grupp}, F. and {Guieu}, S. and {Gullieuszik}, M. and {Haan}, M. de and {Hartke}, J. and {Hartl}, M. and {Haussmann}, F. and {Helin}, T. and {Hess}, H.-J. and {Hofferbert}, R. and {Huber}, H. and {Huby}, E. and {Huet}, J.-M. and {Ives}, D. and {Janssen}, A. and {Jaufmann}, P. and {Jilg}, T. and {Jodlbauer}, D. and {Jost}, J. and {Kausch}, W. and {Kellermann}, H. and {Kerber}, F. and {Kravcar}, H. and {Kravchenko}, K. and {Kulcs{\'a}r}, C. and {Kuncarayakti}, H. and {Kunst}, P. and {Kwast}, S. and {Lang}, F. and {Lange}, J. and {Lapeyrere}, V. and {Le Ruyet}, B. and {Leschinski}, K. and {Locatelli}, H. and {Massari}, D. and {Mattila}, S. and {Mei}, S. and {Merlin}, F. and {Meyer}, E. and {Michel}, C. and {Mohr}, L. and {Montarg{\`e}s}, M. and {M{\"u}ller}, F. and {M{\"u}nch}, N. and {Navarro}, R. and {Neumann}, U. and {Neumayer}, N. and {Neumeier}, L. and {Pedichini}, F. and {Pfl{\"u}ger}, A. and {Piazzesi}, R. and {Pinard}, L. and {Porras}, J. and {Portulari}, E. and {Przybilla}, N. and {Rabien}, S. and {Raffard}, J. and {Ragazzoni}, R. and {Ramlau}, R. and {Ramos}, J. and {Ramsay}, S. and {Raynaud}, H.-F. and {Rhode}, P. and {Richter}, A. and {Rix}, H.-W. and {Rodenhuis}, M. and {Rohloff}, R.-R. and {Romp}, R. and {Rousselot}, P. and {Sabha}, N. and {Sassolas}, B. and {Schlichter}, J. and {Schuil}, M. and {Schweitzer}, M. and {Seemann}, U. and {Sevin}, A. and {Simioni}, M. and {Spallek}, L. and {S{\"o}nmez}, A. and {Suuronen}, J. and {Taburet}, S. and {Thomas}, J. and {Tisserand}, E. and {Vaccari}, P. and {Valenti}, E. and {Verdoes Kleijn}, G. and {Verdugo}, M. and {Vidal}, F. and {Wagner}, R. and {Wegner}, M. and {Winden}, D. van and {Witschel}, J. and {Zanella}, A. and {Zeilinger}, W. and {Ziegleder}, J. and {Ziegler}, B.},
        title = "{The MICADO first light imager for the ELT: overview and current status}",
    booktitle = {Ground-based and Airborne Instrumentation for Astronomy X},
         year = 2024,
       editor = {{Bryant}, Julia J. and {Motohara}, Kentaro and {Vernet}, Jo{\"e}l. R.~D.},
       series = {Society of Photo-Optical Instrumentation Engineers (SPIE) Conference Series},
       volume = {13096},
        month = jul,
          eid = {1309611},
        pages = {1309611},
          doi = {10.1117/12.3017752},
archivePrefix = {arXiv},
       eprint = {2408.16396},
 primaryClass = {astro-ph.IM},
       adsurl = {https://ui.adsabs.harvard.edu/abs/2024SPIE13096E..11S}
}

@ARTICLE{Hammer2021,
       author = {{Hammer}, F. and {Morris}, S. and {Cuby}, J.-G. and {Kaper}, L. and {Steinmetz}, M. and {Afonso}, J. and {Barbuy}, B. and {Bergin}, E. and {Finogenov}, A. and {Gallego}, J. and {Kassin}, S. and {Miller}, C. and {{\"O}stlin}, G. and {Pentericci}, L. and {Schaerer}, D. and {Ziegler}, B. and {Chemla}, F. and {Dalton}, G. and {De Frondat}, F. and {Evans}, C. and {Le Mignant}, D. and {Puech}, M. and {Rodrigues}, M. and {Sanchez-Janssen}, R. and {Taburet}, S. and {Tasca}, L. and {Yang}, Y. and {Zanchetta}, S. and {Dohlen}, K. and {Dubbeldam}, M. and {El Hadi}, K. and {Janssen}, A. and {Kelz}, A. and {Larrieu}, M. and {Lewis}, I. and {MacIntosh}, M. and {Morris}, T. and {Navarro}, R. and {Seifert}, W.},
        title = "{MOSAIC on the ELT: High-multiplex Spectroscopy to Unravel the Physics of Stars and Galaxies from the Dark Ages to the Present Day}",
      journal = {The Messenger},
         year = 2021,
        month = mar,
       volume = {182},
        pages = {33-37},
          doi = {10.18727/0722-6691/5220},
archivePrefix = {arXiv},
       eprint = {2011.03549},
 primaryClass = {astro-ph.GA},
       adsurl = {https://ui.adsabs.harvard.edu/abs/2021Msngr.182...33H}
}

\begin{appendix}
\nolinenumbers 

\section{Details of the model calculations with TURBOSPECTRUM}
\label{app:stellar_models}

The synthetic spectra used by the emulator were pre-computed with
\textsc{Turbospectrum} from a grid of MARCS model atmospheres
\citep{Gustafsson2008}. This appendix summarises several modelling
assumptions and approximations that are not discussed in detail in the main
text.

The atmospheric structures were taken from the public MARCS grid and adopt
the corresponding MARCS solar abundance mixture. For each value of the grid
metallicity, the abundance pattern was scaled globally with $[{\rm M/H}]$.
No independent variations of individual elements, $\alpha$-element abundances,
or CNO abundances were included. The synthetic spectra were computed under
the assumption of local thermodynamic equilibrium (LTE), consistent with
the underlying MARCS atmosphere grid.

A practical limitation of this approach is that the MARCS atmospheric
structures are available only for a discrete set of microturbulent
velocities. In the parameter range relevant for the Cepheid grid, we used
MARCS structures computed for microturbulent velocities of 2 and
5~km~s$^{-1}$. The spectral synthesis itself was performed with
\textsc{Turbospectrum} for microturbulent velocities between
1 and 7~km~s$^{-1}$. For each synthesis calculation we therefore adopted
the MARCS atmosphere with the closest available microturbulence:
the 2~km~s$^{-1}$ MARCS structures were used for
$\xi = 1$, 2, 3, and 4~km~s$^{-1}$, while the
5~km~s$^{-1}$ structures were used for larger values of $\xi$.

This introduces a small inconsistency whenever the microturbulence used in
the atmospheric structure differs from that adopted in the spectral
synthesis. In particular, the turbulent-pressure contribution included in
the MARCS structures is tied to the atmospheric microturbulence rather than
to the value adopted in \textsc{Turbospectrum}. Given the modest
differences involved and the low spectral resolution of the present study,
this effect is expected to be small compared with other modelling
uncertainties. Nevertheless, it remains one of the approximations inherent
to the construction of the grid.

The atomic line list was taken from VALD
\citep{Kupka1999,Ryabchikova2015}, while the molecular data were adopted
from the compilation by B. Plez\footnote{https://www.lupm.in2p3.fr/users/plez/}. 
No empirical recalibration of
oscillator strengths was performed; all transition probabilities were used
as provided by the original sources. This choice keeps the model grid fully
reproducible, but it also means that residual line-list uncertainties are
absorbed by the likelihood model and by the external validation against
optical metallicities, rather than removed through astrophysical tuning of
the synthetic spectra.

The assumption of a scaled-solar abundance pattern is a further limitation
for classical Cepheids. Their surface CNO abundances can be modified by
evolutionary mixing, whereas the present grid changes C, N, and O only
through the global metallicity scaling. Consequently, the strengths of CN
and CO features at fixed metallicity may not be reproduced perfectly by the
models. The impact is expected to be greatest for the coolest stars in the
sample, where molecular opacity contributes more strongly to the
near-infrared spectrum. Possible changes in the atmospheric structure
associated with non-solar CNO mixtures were not explored. Future grids that
allow independent CNO variations would therefore be valuable for assessing
whether molecular-band residuals influence the inferred low-resolution
metallicity scale.

Only the $Y$- and $J$-bands were used for the fits presented in the main
text. The available $K$-band spectra were not used during fitting,
likelihood construction, or parameter inference. Where $K$-band spectra are
available, however, the same best-fitting models provide an independent
qualitative check of the predicted CO band-head region. These comparisons do not contribute to the likelihood and
therefore do not constrain the inferred parameters. Nevertheless, the predicted CO-band behaviour is generally
compatible with the observed $K$-band spectra, suggesting that the adopted
models capture the overall molecular properties of the stars at the level
required for the present $Y+J$ metallicity validation. Examples of these
qualitative $K$-band comparisons are presented in
Appendix~\ref{app:production_parameters}.

\section{Details of the homogenisation of the literature iron abundances}
\label{app:literature_homogenisation}

The external validation presented in this work relies on published optical iron abundances for Galactic Cepheids. Since these measurements originate from multiple studies employing different observational material, analysis techniques, model atmospheres, line lists, and adopted stellar parameters, systematic zero-point differences between literature sources are expected. To minimise these effects and place the comparison on a common footing, we homogenised the literature metallicities onto a single reference scale before comparing them with the SpeX-based metallicities derived in this work.

Throughout this paper, the parameter inferred from the SpeX analysis is the global metallicity $[{\rm M/H}]$, while the external validation is performed against literature optical $[{\rm Fe/H}]$ measurements. The purpose of the homogenisation is therefore not to construct a new Cepheid abundance scale but to provide the most internally consistent optical reference against which the near-infrared metallicities can be evaluated.

We adopted the metallicity scale of \citet{Luck2018} as the reference system. This choice is motivated by the homogeneous reanalysis presented in that work, based on high-resolution optical spectroscopy obtained at multiple pulsation phases for a large sample of Galactic Cepheids. The resulting abundance scale has become widely used in subsequent Cepheid studies and provides one of the most extensive and internally consistent compilations currently available. Recent work by \citet{Bhardwaj2023} also found excellent agreement with the \citet{Luck2018} metallicity scale, further supporting its suitability as the reference system adopted here.

For each literature source included in our comparison sample, we determined a zero-point correction relative to the \citet{Luck2018} scale using stars in common between both studies. For a given literature source $s$, the correction was computed as
\[
\Delta_s
=
{\rm median}
\left(
[\mathrm{Fe/H}]_s
-
[\mathrm{Fe/H}]_{\rm Luck18}
\right),
\]
where the median is evaluated over all stars in common with \citet{Luck2018}. The homogenised abundance adopted in this work is then
\[
[\mathrm{Fe/H}]_{\rm hom}
=
[\mathrm{Fe/H}]_s
-
\Delta_s.
\]
This procedure preserves the relative abundance ranking within each study while reducing systematic offsets between literature scales. The corrections adopted for the individual literature sources are summarised in Table~\ref{tab:lit_homogenisation}, together with the number of stars used to derive each correction.

\begin{table}
\caption{Zero-point corrections to literature iron abundances.}
\label{tab:lit_homogenisation}
\centering
\begin{tabular}{lcc}
\hline
Source & $N_{\rm common}$ & $\Delta_s$ (dex) \\
\hline
\cite{Andrievsky2013}  &  198 & 0.02$\pm$0.06 \\
\cite{Bhardwaj2023}    &   35 & 0.02$\pm$0.10 \\
\cite{Luck2011}        &  280 & 0.09$\pm$0.06 \\ 
\hline
\end{tabular}
\tablefoot{
$N_{\rm common}$ denotes the number of stars in common with \citet{Luck2018} used to determine the median zero-point correction $\Delta_s$.
}
\end{table}

Although the homogenisation procedure reduces the dominant source-to-source zero-point differences, it does not eliminate all systematic effects. Several literature compilations are not fully independent, as they often incorporate overlapping spectroscopic material and previously homogenised abundance measurements. Furthermore, differences in spectral resolution, wavelength coverage, adopted atmospheric parameters, line selection, and abundance methodology can introduce residual systematic differences that are not captured by a simple zero-point correction.

An additional source of scatter arises from the pulsating nature of Cepheids. Although the iron abundance itself is expected to remain constant, abundance determinations obtained from spectra acquired at different pulsation phases can exhibit non-negligible scatter owing to changes in atmospheric structure, line-formation conditions, and analysis systematics throughout the pulsation cycle. The reference abundances of \citet{Luck2018} benefit from extensive multi-phase observations and therefore average over much of this phase-dependent variability.

The homogenised literature metallicities should therefore be regarded as the best available common optical reference scale for the present validation exercise rather than as a noise-free ground truth. Their purpose is to reduce systematic differences across studies and to provide a consistent optical reference for assessing the agreement between the low-resolution near-infrared metallicities derived here and existing optical abundance measurements.

\section{Alternative likelihood formulations and diagnostic tests}
\label{app:likelihood_tests}
\label{app:alternative_likelihoods}

\begin{figure}
\centering
\includegraphics[width=0.98\columnwidth]{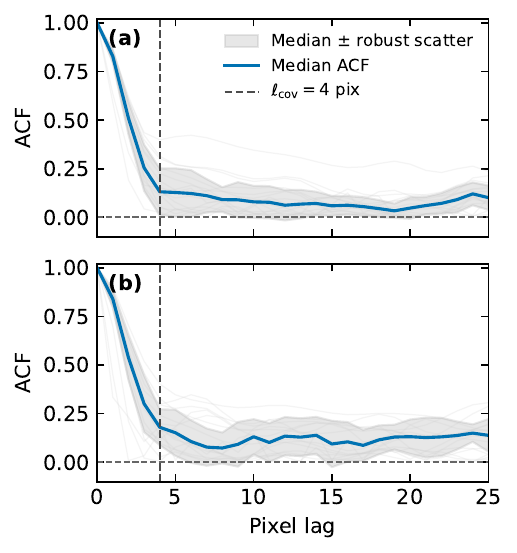}
\caption{Residual autocorrelation functions (ACFs) derived from the
diagonal-likelihood fits for the full sample of 16 Cepheids. Panels
(a) and (b) correspond to the $Y$ and $J$ bands, respectively. Light
grey curves represent the individual stellar ACFs, while the blue
curve denotes the sample median. The shaded region indicates the
median $\pm$ the robust scatter. The dashed vertical line marks a
representative covariance length scale of $\ell_{\rm cov}=4$ pixels.}
\label{fig:acf_length}
\end{figure}

The production likelihood adopted throughout this work is described in Sect.~\ref{sec:likelihood}. During the development of the analysis framework, we explored a broader family of likelihood formulations intended to account for structured model--data residuals beyond the assumption of independent Gaussian noise. Fits obtained with a purely diagonal likelihood exhibit significant residual autocorrelation in both the $Y$ and $J$ bands. Figure~\ref{fig:acf_length} shows the median autocorrelation functions of the residuals together with the star-to-star scatter. Characteristic correlation lengths were estimated from the lag at which the normalised autocorrelation function decreases below $1/e$. The correlations are predominantly short range, with characteristic scales of approximately 3--5 pixels, providing empirical motivation for the introduction of an explicit covariance component in the likelihood and for the exploration of alternative correlated-noise formulations.

The purpose of these experiments was not to identify the most statistically flexible description of the residuals but to assess whether alternative treatments of correlated residual structure could improve the external metallicity validation while preserving the astrophysical information contained in the spectra.
The principal results of these experiments are summarised in Sect.~\ref{sec:likelihood_sensitivity}. The tests described below should therefore be viewed as diagnostic investigations rather than alternative production models. In several cases, increased flexibility improved the formal description of the residuals while simultaneously degrading the agreement with independent metallicity measurements. This behaviour suggests that sufficiently flexible residual models may absorb part of the spectral information that constrains the atmospheric parameters.

\begin{table*}
\caption{Alternative likelihood formulations explored during method development.}
\label{tab:likelihood_summary}
\centering
\begin{tabularx}{\textwidth}{XXXl}
\hline\hline
Likelihood family &
Main idea &
Additional parameters &
Adopted for production? \\
\hline
Diagonal Gaussian &
Independent-pixel noise model &
None &
No \\
Red-noise / wavelet &
Correlated residuals represented through one or more characteristic scales &
Correlation scale(s) &
No \\
Global covariance &
Stationary covariance kernels (squared-exponential or Mat\'ern) &
Amplitude and correlation length &
Yes (regularised form) \\
PCA low-rank residual covariance &
Empirical residual basis derived from principal components &
Mode amplitudes $\lambda_k$ &
No \\
NMF residual diagnostics &
Residual decomposition into non-negative basis functions &
Basis coefficients &
No \\
\hline
\end{tabularx}
\tablefoot{Quantitative performance metrics are provided in Table~\ref{tab:likelihood_sensitivity}.}
\end{table*}

\subsection{Red-noise likelihoods}

The first family of alternatives treated correlated residuals through a red-noise model characterised by one or more correlation scales. The motivation was the residual autocorrelation observed in fits performed with a purely diagonal likelihood. The measured autocorrelation functions typically exhibited characteristic scales of a few pixels, suggesting that neighbouring spectral pixels cannot be regarded as statistically independent.

Several correlation scales were explored, including approximately 2, 4, 8, and 16 pixels, as well as combinations of multiple scales. The dominant contribution was consistently associated with scales of a few pixels, broadly consistent with the residual autocorrelation analysis shown in Fig.~\ref{fig:acf_length}. Although these likelihoods provided a modestly improved description of the residual structure relative to the diagonal likelihood, the resulting metallicities remained broadly 
similar, and no clear advantage emerged relative to the covariance-based formulation adopted for the production analysis.

\subsection{Alternative global covariance models}

We also explored covariance likelihoods based on stationary kernels. These models describe the covariance between pixels $i$ and $j$ through
\[
C_{ij}
=
\sigma_i^2 \delta_{ij}
+
A^2 k(r_{ij}),
\]
where $A$ is the covariance amplitude, $r_{ij}$ denotes the separation between the pixels, and $k(r)$ is a covariance kernel.

The first class of tests employed squared-exponential kernels,
\[
k(r)
=
\exp\!\left(
-\frac{r^2}{2\ell^2}
\right),
\]
while a second class adopted Mat\'ern-3/2 kernels,
\[
k(r)
=
\left(
1+\frac{\sqrt{3}\,r}{\ell}
\right)
\exp\!\left(
-\frac{\sqrt{3}\,r}{\ell}
\right),
\]
where $\ell$ is the characteristic correlation length. The Mat\'ern formulation permits less smooth residual behaviour and generally provided a more realistic description of the observed residual structure.

In addition to the kernel shape itself, we investigated different prior assumptions for the covariance amplitude. Broad, weakly informative priors frequently drove the posterior towards large covariance amplitudes. This behaviour increased the formal parameter uncertainties and improved some uncertainty-calibration metrics, but it also reduced the constraining power of the spectra and increased the scatter of the inferred metallicities relative to the external reference scale. These experiments motivated the adoption of informative truncated priors in the production likelihood. The final implementation should therefore be regarded as a regularised covariance model rather than as a fully data-driven covariance inference.

\subsection{Low-rank residual covariance from principal-component analysis}

The most ambitious residual model explored in this work employed empirical residual bases derived from principal-component analysis (PCA; \citealt{Jolliffe2002}). The underlying idea was that at least part of the residual structure may arise from repeatable patterns shared among different stars, such as systematic molecular-band mismatches, imperfect line strengths, telluric-correction residuals, continuum-placement effects, or deficiencies in the synthetic spectra. If such structures can be represented by a small number of basis vectors, they may be incorporated through a low-rank covariance contribution.

The analysis began by constructing a residual matrix from the baseline fits,
\[
R(\lambda,s)
=
\frac{
f_{\rm obs}(\lambda,s)
-
f_{\rm mod}(\lambda,s)
}
{
\sigma(\lambda,s)
},
\]
where $s$ denotes the star index. The residuals were normalised by their uncertainties before the PCA decomposition. The resulting eigenvectors are therefore dimensionless and describe patterns in units of the local noise.

A subtle but important consequence is that the corresponding covariance contribution must be constructed in flux units,
\[
C_{{\rm LR},ij}
=
\sigma_i \sigma_j
\sum_{k=1}^{K}
\lambda_k^2
e_{ik}
e_{jk},
\]
where $e_{ik}$ are the PCA eigenvectors and $\lambda_k$ are the amplitudes of the retained modes. In some exploratory implementations, the covariance contribution was additionally restricted to selected wavelength regions through multiplicative masks applied to the eigenvectors.

The PCA approach proved informative but also revealed important limitations. Principal components identify directions of maximum variance rather than physically meaningful residual structures. For the present data set, the PCA basis was derived from residuals of only 16 stars, limiting the ability of the decomposition to separate distinct physical sources of model mismatch. The leading components are therefore expected to combine several effects, including molecular-band residuals, imperfect line strengths, telluric residuals, and continuum mismatches, and do not admit a straightforward physical interpretation.

More importantly, the low-rank models consistently degraded the external metallicity validation relative to the regularised covariance likelihood, yielding larger metallicity scatter and poorer uncertainty calibration. This behaviour suggests that the PCA basis vectors may absorb part of the astrophysical information that constrains the stellar parameters. Although the approach remains attractive in principle, especially for substantially larger training samples, we do not consider the present data set sufficiently large to support a reliable empirical decomposition of the residual structure.

\subsection{Non-negative matrix factorisation residual bases}

As an alternative to PCA, we investigated residual decompositions based on non-negative matrix factorisation (NMF; \citealt{Lee1999,Lee2001}). Rather than decomposing the signed residuals themselves, NMF was primarily explored as a tool for identifying regions that contribute disproportionately to the residual variance. Compared to PCA, NMF was viewed primarily as a method for identifying problematic wavelength regions rather than for constructing a generative residual model.

The residual matrix can be approximated as
\[
R \approx A E,
\]
where the matrices $A$ and $E$ contain non-negative basis vectors and coefficients.

Unlike PCA, NMF does not rely on positive and negative cancellations and often produces more localised basis functions. In practice, the NMF decomposition proved useful for identifying wavelength regions associated with persistent residual structures and therefore served as a diagnostic tool for assessing model shortcomings. However, incorporating these components directly into the likelihood did not improve the metallicity validation relative to the adopted covariance model. Consequently, NMF was retained only as an exploratory diagnostic.

\subsection{Implications for the adopted likelihood}

Taken together, the experiments described above illustrate an important trade-off between residual flexibility and parameter identifiability. The diagonal likelihood underestimates the impact of correlated residual structure, whereas highly flexible empirical residual models can absorb part of the astrophysical information contained in the spectra. The regularised covariance likelihood adopted for the production analysis was therefore selected as a balance between these two extremes.

The final model explicitly accounts for short-range residual correlations and localised problematic spectral regions while remaining sufficiently constrained to preserve the metallicity information encoded in the low-resolution $Y$+$J$ spectra. In the context of the present validation study, this balance proved more important than maximising the formal flexibility of the residual model.

\section{Complete posterior parameters and supplementary fit diagnostics}
\label{app:production_parameters}

This appendix summarises the posterior parameter estimates obtained with the production likelihood and provides supplementary figures illustrating the quality of the final spectral fits.

\begin{sidewaysfigure*}
\centering
\includegraphics{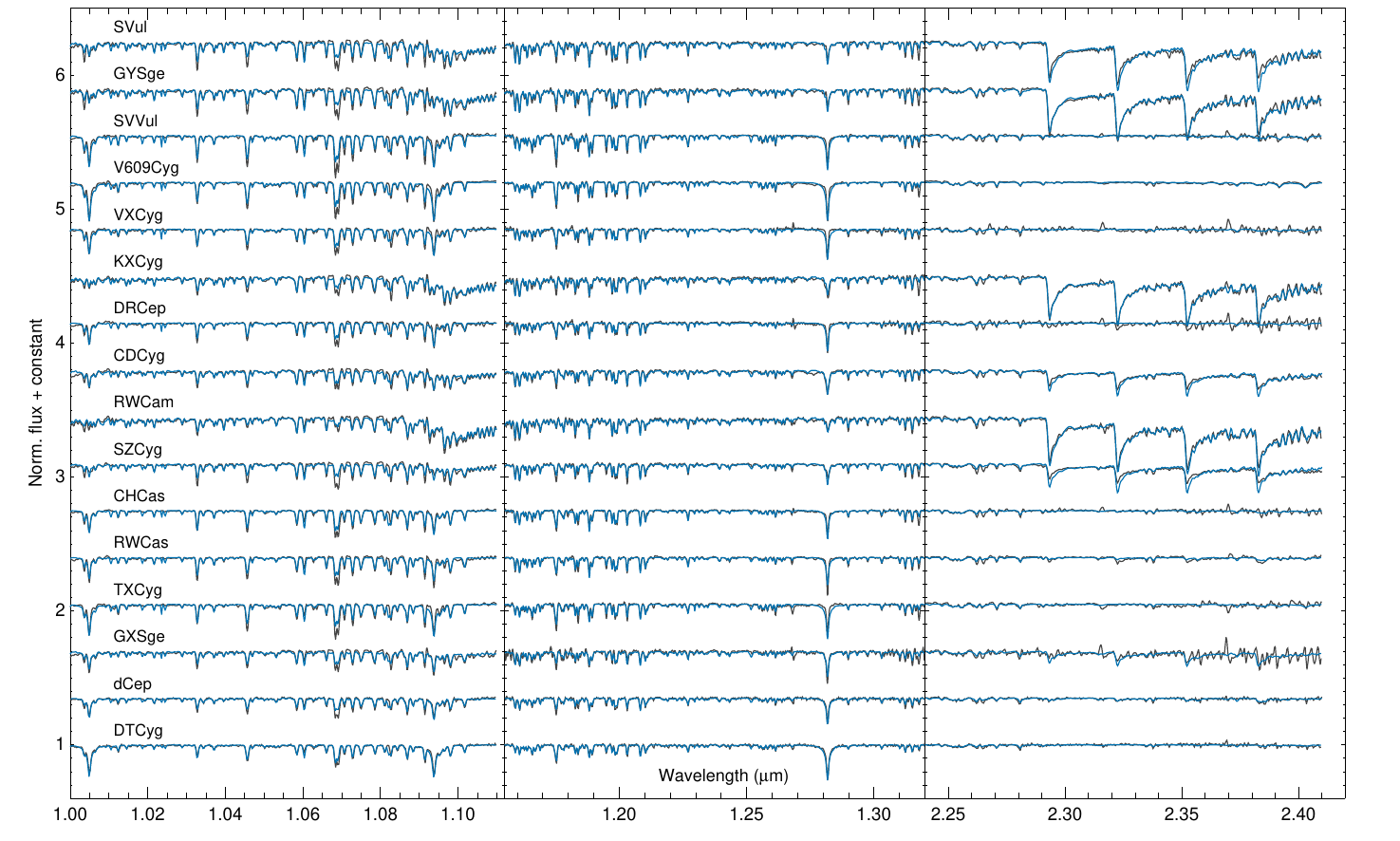}
\caption{Global comparison between the observed spectra and the final
best-fitting models for the full Cepheid sample. The three panels show
the $Y$, $J$, and $K$ bands, respectively. The $Y$- and $J$-band
regions were included in the inference, while the $K$-band models are
posterior predictions outside the fitted wavelength range. Individual
spectra have been vertically shifted for clarity and are ordered from
top to bottom by decreasing pulsation period. Black lines represent
the observed spectra, and blue lines the corresponding model spectra.}
\label{fig:global_fits}
\end{sidewaysfigure*}

Figure~\ref{fig:global_fits} presents a compact overview of the observed spectra and corresponding best-fitting models for the complete sample. The $Y$- and $J$-band panels show the wavelength regions used during the inference, while the $K$-band panel presents posterior predictions outside the fitted wavelength range. The figure provides a compact visual summary of the overall quality of the fits and of the general consistency of the posterior predictions outside the fitted wavelength range.

\begin{figure*}
\centering
\includegraphics[width=0.8\textwidth]{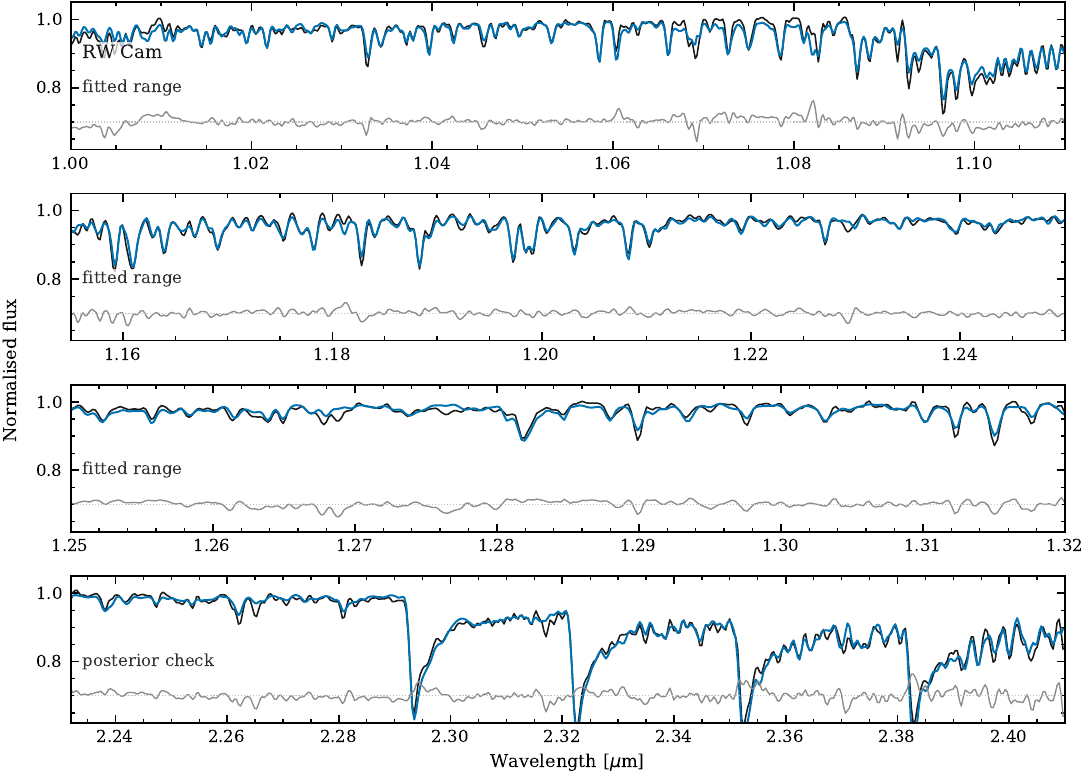}
\caption{Representative spectral fit for RW~Cam. The observed SpeX
spectrum is shown in black, the best-fitting model in blue, and the
residuals in grey, shifted downward for clarity. The upper three panels
show the $Y$- and $J$-band wavelength intervals included in the
likelihood. The lower panel shows the corresponding $K$-band posterior
prediction, which was not included in the fit and serves only as a
qualitative consistency check.}
\label{fig:fit_rwcam}
\end{figure*}

\begin{figure*}
\centering
\includegraphics[width=0.8\textwidth]{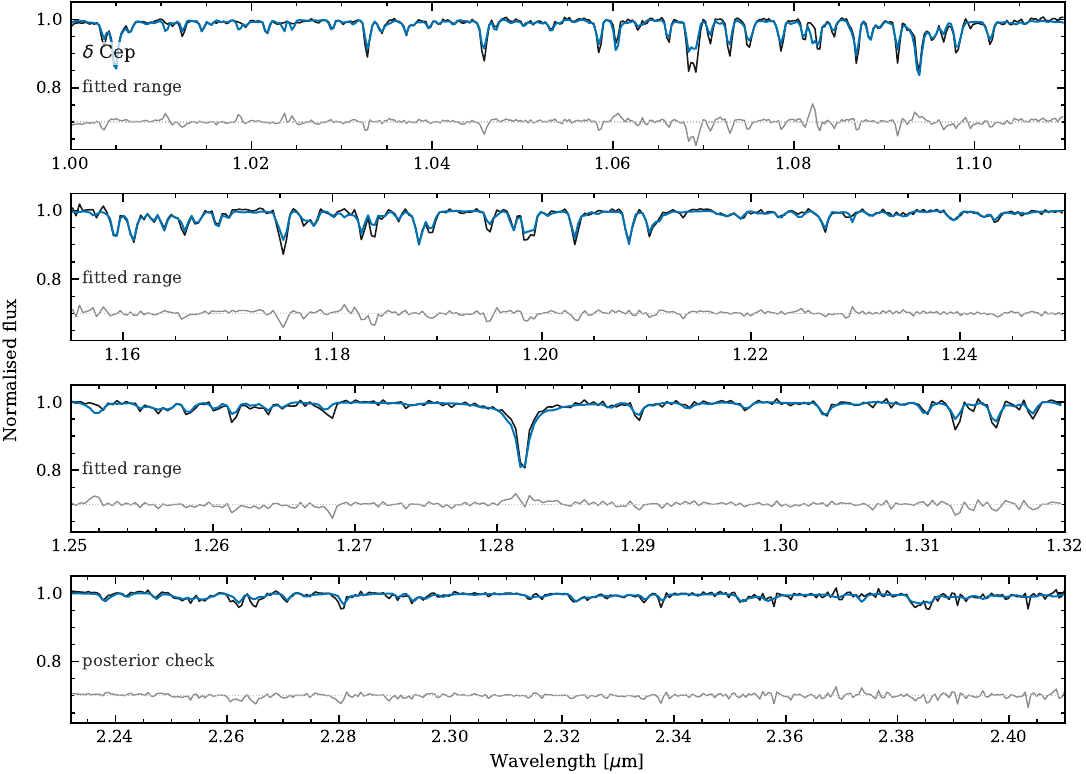}
\caption{Representative spectral fit for $\delta$~Cep. The layout and
conventions are the same as in Fig.~\ref{fig:fit_rwcam}.
}
\label{fig:fit_dcep}
\end{figure*}

Figures~\ref{fig:fit_rwcam} and \ref{fig:fit_dcep} show two representative examples in greater detail. In each case, the lower panel presents the corresponding $K$-band prediction and serves as an independent qualitative consistency check.

Table~\ref{tab:posterior_parameters} lists the posterior median values and central 68\% credible intervals for all fitted parameters, including the atmospheric parameters, the resolving-power scale factor, and the residual-model hyperparameters.

\begin{table*}
\caption{Posterior parameter estimates obtained with the production likelihood.}
\label{tab:posterior_parameters}
\centering
\begin{tabular}{lcccccccc}
\hline\hline
ID &
$T_{\rm eff}$ &
$\log g$ &
$\xi$ &
$[{\rm M/H}]$ &
$s_R$ &
$\log s_{\rm cov}$ &
$\ell_{\rm cov}$ &
$\log a_{\rm flag}$ 
\\
&
(K) &
(dex) &
(km\,s$^{-1}$) &
(dex) &
&
&
(pix) &
 \\
\hline

S Vul    & $5110^{+134}_{-122}$ & $0.71^{+0.17}_{-0.14}$ & $4.3^{+0.9}_{-0.8}$ & $0.06^{+0.18}_{-0.20}$ & $1.02^{+0.06}_{-0.07}$ & $-1.00^{+0.05}_{-0.04}$ & $4.01^{+0.36}_{-0.38}$ & $0.50^{+0.06}_{-0.04}$ \\
GY Sge   & $5130^{+119}_{-136}$ & $0.81^{+0.16}_{-0.15}$ & $4.4^{+0.8}_{-0.9}$ & $0.15^{+0.18}_{-0.17}$ & $1.01^{+0.06}_{-0.07}$ & $-1.00^{+0.05}_{-0.05}$ & $4.01^{+0.36}_{-0.39}$ & $0.50^{+0.06}_{-0.04}$ \\
SV Vul   & $6238^{+226}_{-175}$ & $1.39^{+0.17}_{-0.15}$ & $6.8^{+1.2}_{-0.2}$ & $0.05^{+0.15}_{-0.16}$ & $1.01^{+0.06}_{-0.07}$ & $-1.00^{+0.06}_{-0.04}$ & $4.00^{+0.36}_{-0.40}$ & $0.51^{+0.06}_{-0.04}$ \\
V609 Cyg & $6469^{+233}_{-101}$ & $1.63^{+0.16}_{-0.17}$ & $6.4^{+1.6}_{-0.3}$ & $-0.02^{+0.17}_{-0.18}$ & $1.01^{+0.07}_{-0.07}$ & $-1.00^{+0.06}_{-0.04}$ & $3.99^{+0.43}_{-0.36}$ & $0.50^{+0.06}_{-0.04}$ \\
VX Cyg   & $5916^{+225}_{-214}$ & $1.54^{+0.16}_{-0.19}$ & $4.8^{+2.6}_{-1.1}$ & $-0.18^{+0.25}_{-0.28}$ & $1.01^{+0.07}_{-0.07}$ & $-1.00^{+0.06}_{-0.05}$ & $3.99^{+0.38}_{-0.37}$ & $0.50^{+0.06}_{-0.04}$  \\
KX Cyg   & $5003^{+182}_{-156}$ & $1.06^{+0.17}_{-0.16}$ & $2.5^{+0.8}_{-0.8}$ & $0.14^{+0.24}_{-0.22}$ & $1.01^{+0.06}_{-0.07}$ & $-1.01^{+0.05}_{-0.05}$ & $3.99^{+0.39}_{-0.34}$ & $0.50^{+0.06}_{-0.04}$ \\
DR Cep   & $5702^{+216}_{-247}$ & $1.45^{+0.18}_{-0.20}$ & $5.2^{+2.2}_{-1.2}$ & $-0.32^{+0.20}_{-0.27}$ & $1.01^{+0.06}_{-0.08}$ & $-1.00^{+0.05}_{-0.05}$ & $4.01^{+0.36}_{-0.38}$ & $0.49^{+0.06}_{-0.05}$ \\
CD Cyg   & $5456^{+181}_{-211}$ & $1.37^{+0.20}_{-0.17}$ & $2.8^{+0.8}_{-1.6}$ & $0.18^{+0.30}_{-0.22}$ & $1.01^{+0.07}_{-0.07}$ & $-1.00^{+0.05}_{-0.05}$ & $3.99^{+0.39}_{-0.35}$ & $0.50^{+0.07}_{-0.04}$ \\
RW Cam   & $4634^{+151}_{-144}$ & $0.90^{+0.18}_{-0.15}$ & $2.8^{+0.7}_{-0.7}$ & $-0.07^{+0.20}_{-0.22}$ & $1.02^{+0.07}_{-0.06}$ & $-1.00^{+0.06}_{-0.05}$ & $4.00^{+0.38}_{-0.38}$ & $0.49^{+0.06}_{-0.04}$ \\
SZ Cyg   & $5322^{+180}_{-177}$ & $1.34^{+0.17}_{-0.17}$ & $3.9^{+1.2}_{-1.8}$ & $0.04^{+0.24}_{-0.28}$ & $1.00^{+0.07}_{-0.06}$ & $-1.00^{+0.06}_{-0.04}$ & $4.02^{+0.35}_{-0.41}$ & $0.50^{+0.06}_{-0.04}$  \\
CH Cas   & $5842^{+213}_{-209}$ & $1.60^{+0.18}_{-0.17}$ & $5.8^{+2.1}_{-0.5}$ & $-0.17^{+0.19}_{-0.20}$ & $1.01^{+0.06}_{-0.08}$ & $-1.01^{+0.05}_{-0.05}$ & $4.02^{+0.35}_{-0.37}$ & $0.49^{+0.06}_{-0.04}$ \\
RW Cas   & $5730^{+207}_{-208}$ & $1.56^{+0.16}_{-0.20}$ & $5.4^{+1.9}_{-1.1}$ & $-0.08^{+0.20}_{-0.24}$ & $1.01^{+0.06}_{-0.08}$ & $-1.01^{+0.05}_{-0.05}$ & $3.99^{+0.33}_{-0.44}$ & $0.50^{+0.06}_{-0.04}$ \\
TX Cyg   & $6266^{+218}_{-199}$ & $1.78^{+0.17}_{-0.18}$ & $6.0^{+2.0}_{-0.4}$ & $-0.03^{+0.18}_{-0.20}$ & $1.02^{+0.06}_{-0.08}$ & $-1.00^{+0.05}_{-0.04}$ & $4.01^{+0.36}_{-0.38}$ & $0.50^{+0.06}_{-0.04}$ \\
GX Sge   & $5574^{+213}_{-230}$ & $1.52^{+0.18}_{-0.19}$ & $2.7^{+0.7}_{-1.7}$ & $0.16^{+0.30}_{-0.24}$ & $1.01^{+0.06}_{-0.08}$ & $-1.00^{+0.06}_{-0.04}$ & $4.00^{+0.36}_{-0.40}$ & $0.50^{+0.06}_{-0.04}$ \\
$\delta$ Cep & $5800^{+207}_{-226}$ & $1.91^{+0.18}_{-0.19}$ & $3.9^{+1.7}_{-2.0}$ & $-0.03^{+0.30}_{-0.25}$ & $1.01^{+0.08}_{-0.07}$ & $-1.01^{+0.05}_{-0.05}$ & $4.01^{+0.38}_{-0.37}$ & $0.49^{+0.06}_{-0.05}$  \\
DT Cyg   & $6356^{+238}_{-182}$ & $2.44^{+0.17}_{-0.17}$ & $4.5^{+2.9}_{-1.1}$ & $-0.14^{+0.28}_{-0.28}$ & $1.00^{+0.08}_{-0.07}$ & $-1.00^{+0.05}_{-0.05}$ & $4.01^{+0.36}_{-0.39}$ & $0.49^{+0.06}_{-0.04}$ \\
\hline
\end{tabular}
\tablefoot{Values are posterior medians with lower and upper uncertainties. The covariance length $\ell_{\rm cov}$ is sampled in pixels and internally converted to a local velocity scale, as described in Sect.~\ref{subsec:covariance_likelihood}.}
\end{table*}

\section{$J$-band-only analysis}
\label{app:jband}

The analysis presented throughout this paper is based on the combined $Y$+$J$ wavelength intervals. This choice is motivated by the broad wavelength coverage provided by instruments such as SpeX and by future facilities capable of obtaining both bands simultaneously. However, there are important observational situations in which only a single near-infrared band may be available. Examples include multi-object spectrographs such as MOSFIRE at Keck, as well as observing programmes with facilities such as JWST/NIRSpec where only a subset of the near-infrared wavelength range may be available or practical to observe.

A further consideration is that Cepheids are variable stars. When different wavelength bands are obtained at different epochs, changes in the atmospheric structure over the pulsation cycle may introduce systematic differences between the spectra. In such situations, a single-band analysis may in some cases be preferable to combining non-simultaneous observations obtained at different phases.

To investigate this question, we repeated the full production analysis described in Sect.~\ref{sec:likelihood} using the same spectral models, priors, covariance treatment, and inference procedure, but retaining only the $J$-band fitting windows. No other aspect of the methodology was modified. The purpose of this experiment is therefore not to optimise a dedicated $J$-band analysis, but to quantify the impact of the reduced wavelength coverage under otherwise identical assumptions.

We focus on the $J$-band because, at the resolution considered here, it contains several of the strongest metal-line blends available in the near-infrared and has been widely used for abundance work on cool luminous stars \citep[e.g.][]{Davies2010, Gazak2014, Inno2019}. By contrast, the low-resolution $Y$-band contains fewer prominent metallic features and was therefore not investigated separately.

Table~\ref{tab:jband_comparison} summarises the resulting validation statistics and compares them with those obtained from the adopted $Y$+$J$ analysis. The $J$-band-only analysis produces a noticeably larger systematic offset relative to the homogenised optical scale, with the mean residual changing from $-0.06$\,dex for the combined $Y$+$J$ analysis to $-0.12$\,dex for the $J$-band-only case. The empirical scatter remains comparable, changing only from $0.09$\,dex to $0.08$\,dex, while the median posterior uncertainty increases from $0.23$\,dex to $0.26$\,dex. The uncertainty calibration also degrades, with RMS$(z)$ increasing from $0.41$ to $0.52$.

The $J$-band-only fits nevertheless recover a metallicity scale that remains broadly compatible with the homogenised optical reference abundances. Despite the larger systematic offset and somewhat poorer uncertainty calibration, the similar empirical scatter suggests that the $J$-band retains much of the metallicity information available at these spectral resolutions. The inferred stellar parameters are also generally consistent with those obtained from the full $Y$+$J$ analysis. Effective temperatures, gravities, and microturbulent velocities typically agree within their quoted posterior uncertainties, indicating that the overall parameter constraints remain broadly stable despite the reduced wavelength coverage.

\begin{table}
\caption{Comparison of the combined $Y+J$-band and $J$-band-only
analyses using the production likelihood.}
\label{tab:jband_comparison}
\centering
\begin{tabular}{lcc}
\hline
Metric & $Y+J$ & $J$ only \\
\hline
$\langle\Delta\rangle$ (dex)         & $-0.06$ & $-0.12$ \\
$\sigma_{\Delta}$ (dex)              & $0.09$  & $0.08$ \\
$\sigma_{\rm MAD}$ (dex)             & $0.09$  & $0.09$ \\
Median $\sigma_{[{\rm M/H}]}$ (dex)  & $0.23$  & $0.26$ \\
$\langle z\rangle$                   & $-0.26$ & $-0.44$ \\
RMS$(z)$                             & $0.41$  & $0.52$ \\
\hline
\end{tabular}
\end{table}

These results indicate that useful Cepheid metallicities can still be recovered from low-resolution $J$-band spectroscopy alone. Nevertheless, the combined $Y$+$J$ analysis provides a metallicity scale that is more closely tied to the optical reference abundances and yields slightly tighter parameter constraints. We therefore regard the $J$-band-only results as encouraging for applications where only a single near-infrared band is available, while the simultaneous $Y$+$J$ analysis remains the preferred configuration whenever both bands can be observed together.

More generally, the comparison suggests that broader wavelength coverage improves both the robustness of the inferred parameters and the external calibration of the metallicity scale, providing additional support for the adoption of the combined $Y$+$J$ configuration throughout this work.

\end{appendix}
\end{document}